\documentclass[prl,twocolumn,superscriptaddress]{revtex4-2}

\usepackage{amsmath,amssymb,mathrsfs}
\usepackage{graphicx}
\usepackage{colortbl}
\usepackage{braket}
\usepackage{bm}
\usepackage{amsfonts}
\usepackage{ulem}

\begin{document}

\title{Detecting Many-Body Scars from Fisher Zeros}

\author{Yuchen Meng}
\affiliation {Key Laboratory of Polar Materials and Devices (MOE), School of Physics and Electronic Science, East China Normal University, Shanghai 200241, China}

\author{Songtai Lv}
\affiliation{Key Laboratory of Polar Materials and Devices (MOE), School of Physics and Electronic Science, East China Normal University, Shanghai 200241, China}

\author{Yang Liu}
\affiliation{Key Laboratory of Polar Materials and Devices (MOE), School of Physics and Electronic Science, East China Normal University, Shanghai 200241, China}

\author{Zefan Tan}
\affiliation{Key Laboratory of Polar Materials and Devices (MOE), School of Physics and Electronic Science, East China Normal University, Shanghai 200241, China}

\author{Erhai Zhao}
\altaffiliation{ezhao2@gmu.edu}
\affiliation{Department of Physics and Astronomy, George Mason University, Fairfax, Virginia 22030, USA}

\author{Haiyuan Zou}
\altaffiliation{hyzou@phy.ecnu.edu.cn}
\affiliation{Key Laboratory of Polar Materials and Devices (MOE), School of Physics and Electronic Science, East China Normal University, Shanghai 200241, China}

\begin{abstract}
The far-from-equilibrium dynamics of certain interacting quantum systems still defy precise understanding. One example is the so-called quantum many-body scars (QMBSs), where a set of energy eigenstates evade thermalization to give rise to long-lived oscillations. Despite the success of viewing scars from the perspectives of symmetry, commutant algebra, and quasiparticles, it remains a challenge to elucidate the mechanism underlying all QMBS and to distinguish them from other forms of ergodicity breaking. In this work, we introduce an alternative route to detect and diagnose QMBS based on Fisher zeros, i.e., the patterns of zeros of the analytically continued partition function $Z$ on the complex $\beta$ (inverse temperature) plane. For systems with scars, a continuous line of Fisher zeros will appear off the imaginary $\beta$ axis and extend upward, separating the $\beta$ plane into regions with distinctive thermalization behaviors. This conjecture is motivated from interpreting the complex $Z$ as the return amplitude of the thermofield double state, and it is validated by analyzing two models with QMBS, the $\bar{P}X\bar{P}$ model and the Ising chain in external fields. These models also illustrate the key difference between QMBS and strong ergodicity breaking including their distinctive renormalization group flows on the complex $\beta$ plane. This ``statistical mechanics" approach places QMBS within the same framework of thermal and dynamical phase transitions. It has the advantage of spotting scars without exhaustively examining each individual quantum state.

\end{abstract}

\maketitle

{\it Introduction—}The eigenstate thermalization hypothesis (ETH) postulates that an isolated interacting quantum system typically thermalizes~\cite{Deutsch1991PRA,Srednicki1994PRE}. But there are known cases of ETH violation, even in nonintegrable systems free of disorder. A prominent example is quantum many-body scars (QMBSs)~\cite{Papic2018NP,Bernevig2018PRB,Schecter2019PRL,Robinson2019PRL,Pollmann2020PRX}, which have been observed on various platforms~\cite{Lukin2017Nature,Lukin2021Science,Zhang2022NP,Pan2023PRR_QMBS}. Roughly speaking, a set of energy eigenstates, usually equally spaced to form a tower structure, become decoupled from the rest of the spectrum and give rise to persistent oscillations. The mechanism by which QMBS evade ETH has been attributed to Krylov-restricted Hilbert space fragmentation, spectrum-generating algebras, commutant algebras, or projector embeddings~\cite{QMBSreview2021NP,Moudgalya2022QMBSreview,Moudgalya2024PRX,Shiraishi2017Embedding}, and they can also be understood from a quasiparticle perspective~\cite{QMBS2023ARCMP}. A unifying theoretical framework for all known QMBS, however, is still lacking.

Key properties of QMBS are typically explored by examining the candidate Hamiltonian including its energy spectrum, eigenstates, the entanglement entropy, and Loschmidt echo. As a result, to find new instances of scar states, one must hand select a few states and monitor their dynamics. The exhaustive search can turn prohibitively expensive for large system sizes, evocative of finding a needle in a haystack. Thus it is desirable to develop a diagnostic device that can spot the existence of QMBS while avoiding the inspection of each individual state.
Compared to QMBS, strong breaking of ETH (SBETH) features more profound deviations from thermal equilibrium~\cite{Banuls2011PRL}. Examples include oscillations described by quasiparticles~\cite{Kormos2016NP,Motrunich2017PRA,Collura2018PRA} and continuous time crystals~\cite{Wilczek2012QTC,Nayak2017PRX,Buca2020prb,Hemmerich2022Science,Liu2023NP,Greilich2024NP,You2024NP}, which spontaneously break time-translation symmetry.
Possible connection between QMBS and SBETH has been suggested~\cite{CTCandQMBS2022,Buca2023prx}. But their coexistence and interplay remain poorly understood.

Motivated by these challenges, in this Letter, we propose and demonstrate 
an alternative approach to QMBS. It is based on Fisher zeros originally invented to understand the singularities of equilibrium partition functions and thermal phase transitions (TPTs)~\cite{LeeYang1,LeeYang2,Fisher1965statistical}. The dynamical decoupling of scars from the rest of the spectrum is reminiscent of phase separation and hints that the tools from statistical mechanics may help yield fresh insights. To formalize this idea, we follow Fisher and analytically continue the partition function $Z=\sum_n e^{-\beta E_n}$ by allowing $\beta$ to take on complex values, $\beta=\beta_r + i\beta_i$. Then $Z$ becomes complex valued, $Z=\sum_n e^{-\beta_r E_n} e^{-itE_n}$, which describes the time evolution of a thermal or mixed state at inverse temperature $\beta_r$ and at time $t=\beta_i$. The modulus square of $Z$, known as the spectral form factor, has been widely used in the study of quantum chaos~\cite{Maldacena2016jhep}. To further connect $Z$ to pure state dynamics, we can purify the mixed state by enlarging the Hilbert space, e.g., by making a copy $R$ of the original system $L$ and defining the thermofield double (TFD) state $|\Psi(\beta_r,0)\rangle = \sum_n e^{-\beta_r E_n/2}|n\rangle_L\otimes|n\rangle_R$~\cite{takahashi1996thermo} (the thermal state is obtained by partial trace of $\Psi$ over subsystem $R$). Then $Z$ is nothing but the return amplitude of the pure TFD state evolving under the single Hamiltonian $H_L\otimes 1_R$, $Z(\beta_r,t)\sim \langle\Psi(\beta_r,0)|\Psi(\beta_r,t)\rangle$, where the normalization factor is dropped for brevity and $\beta_r$ serves as a parameter of $\Psi$ that controls the weights of various eigenstates $| n\rangle$~\cite{delCampo2017PRD}. 

The scar states contained in system $L$, if any, will continue to evade thermalization in the doubled system, for the added subsystem $R$ acts merely as a bath. Thus, the existence of scars will manifest in the return amplitude $Z(\beta_r,t)$ at long times if the initial state $|\Psi (\beta_r,0)\rangle$ has sufficient overlap (dictated by amplitudes $\propto e^{-\beta_rE_n/2}$) with the scar states. In other words, $Z$ cannot be a smooth single-valued function free of qualitative changes when $\beta_r$ is varied across the complex $\beta$ plane. 

\begin{figure}[b]
    \includegraphics[width=0.48\textwidth]{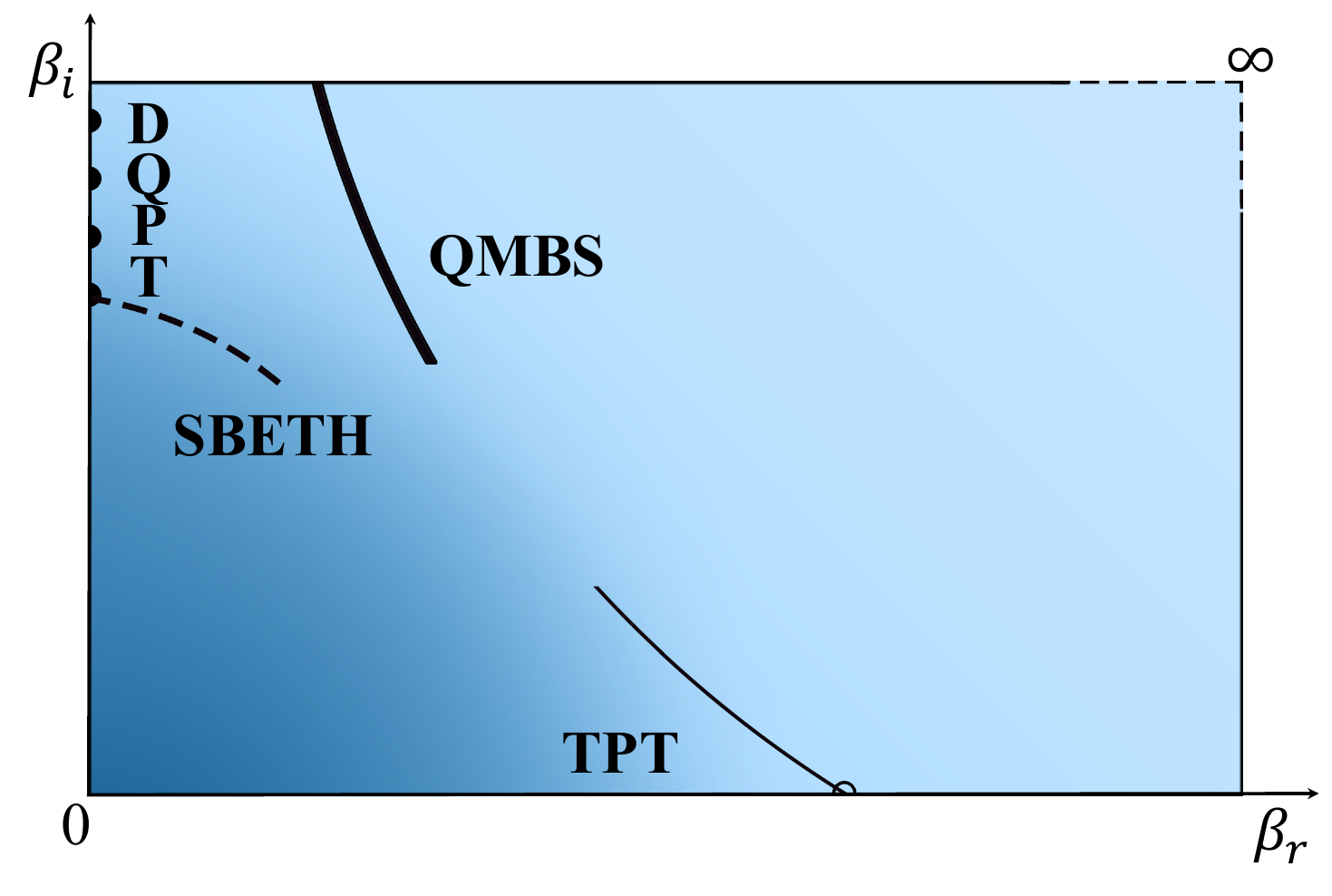}
    \caption{(Schematic) A synopsis of possible locations of Fisher zeros, the zeros of $Z=\mathrm{Tr} e^{-\beta H}$ on the complex $\beta=\beta_r+i\beta_i$ plane in the thermodynamic limit. The thin line illustrates typical Fisher zeros in classical models, its intersection with the $\beta_r$ axis (the empty circle) marks the TPT point. Solid circles on the $\beta_i$ axis mark the DQPT points. The 
dashed line illustrates one scenario of SBETH with the Fisher zeros touching the $\beta_i$ axis. The thick line of zeros off the $\beta_i$ axis extending upward indicates QMBS. Darker regions feature larger thermodynamic and smaller quantum fluctuations.
}
    \label{fig:fig1}
\end{figure}

{\it A conjecture regarding scars and Fisher zeros—}These considerations lead to our main conjecture: for a system with QMBS, there exists at least a line of Fisher zeros that lies off the $\beta_i$ axis and extends to large $\beta_i$ (the long-time limit). As illustrated in Fig.~\ref{fig:fig1}, the line of zeros (labeled by QMBS) extends upward 
and marks the singularity of $f\equiv\ln |Z|$ in the limit of large system sizes. When the line is crossed, e.g., by increasing $\beta_r$ in the horizontal direction, the long-time dynamics of the TFD state will experience a qualitative change. Before we present numerical evidence to support this conjecture, it is illuminating to compare the location and implication of Fisher zeros in the present setting to their previous applications. In his seminar work on classical statistical models, Fisher observed that the zeros of $Z(\beta_r,\beta_i)$ pinching the $\beta_r$ axis marks a TPT. In recent years, the concept of the complex-valued partition function and its zeros has been extended to the field of quantum many-body physics. For example, we have shown that
the vanishing of Fisher zero lines at infinite $\beta$ identifies the quantum critical point of the one-dimensional transverse-field Ising model~\cite{Liu2023CPL,Liu2024PRR,Liu2024CPL}. In quench dynamics, following an initial state $\psi_i$, one can define a boundary partition function $Z_i(\beta_r,\beta_i)=\langle\psi_i|e^{-(\beta_r+i\beta_i)H}|\psi_i\rangle$. The zeros of $Z_i$ (or the singularity of $f_i\equiv\ln|Z_i|$) on the $\beta_i$ axis then mark dynamical quantum phase transitions (DQPTs)~\cite{Heyl2013PRL,reviewDPT2016,Heyl2018review,Halimeh2023prr}. The empty and solid circles in Fig.~\ref{fig:fig1} identify TPT and DQPT, respectively. In contrast, the zeros associated with QMBS are located away from the real and the imaginary $\beta$ axes. To our knowledge, the link between these Fisher zeros and breakdown of ETH has not been recognized before. 

We emphasize that the pattern of Fisher zeros on the complex $\beta$ plane is model dependent and may appear more complicated than the simple schematic in Fig.~\ref{fig:fig1}. In particular, it may contain additional features associated with SBETH~\cite{CTCandQMBS2022} coexisting with QMBS. A typical scenario is shown in Fig.~\ref{fig:fig1}, where lines of Fisher zeros labeled by SBETH repeatedly intersect the $\beta_i$ axis, so $Z$ exhibits more pronounced non-analytical behaviors near the $\beta_i$ axis. To validate the conjecture regarding QMBS, we first construct a generalized $PXP$ model to incorporate a crossover from QMBS to SBETH. Then, a nonintegrable model with Ising interactions is used to explore the rich interplays of the Fisher zeros when both QMBS and SBETH are present.

{\it The $\bar{P}X\bar{P}$ model—}We generalize 
$PXP$-like models~\cite{Papic2018NP,Lesanovsky2012pra,Turner2018prb,Ho2019prl,Deng2023prl,Ivanov2025arxiv}, which describe atoms in the Rydberg-blockade regime, to the following $\bar{P}X\bar{P}$ model, 
\begin{equation}
   {H} = -\sum_{i=1}^{L}\bar{P_i}\sigma^x_{i+1}\bar{P}_{i+2}.
    \label{eq:HamPXP}
\end{equation}
Here $\sigma^x_i$ is the Pauli $x$ matrix operators on site $i$, and periodic boundary condition (PBC) is assumed.  
The operator 
$\bar{P} =  (|0\rangle\langle0|+g|1\rangle\langle1|)$, where $|0\rangle$ and $|1\rangle$ are two eigenstates of $\sigma^z$ corresponding to 0 and 1 atom at a given site, respectively. 
The tuning parameter $g\in [0,1]$ allows for
extrapolation between the $PXP$ model ($g=0$) and the non-interacting single spin flip model ($g=1$).

\begin{figure}[t]
    \includegraphics[width=0.49\textwidth]{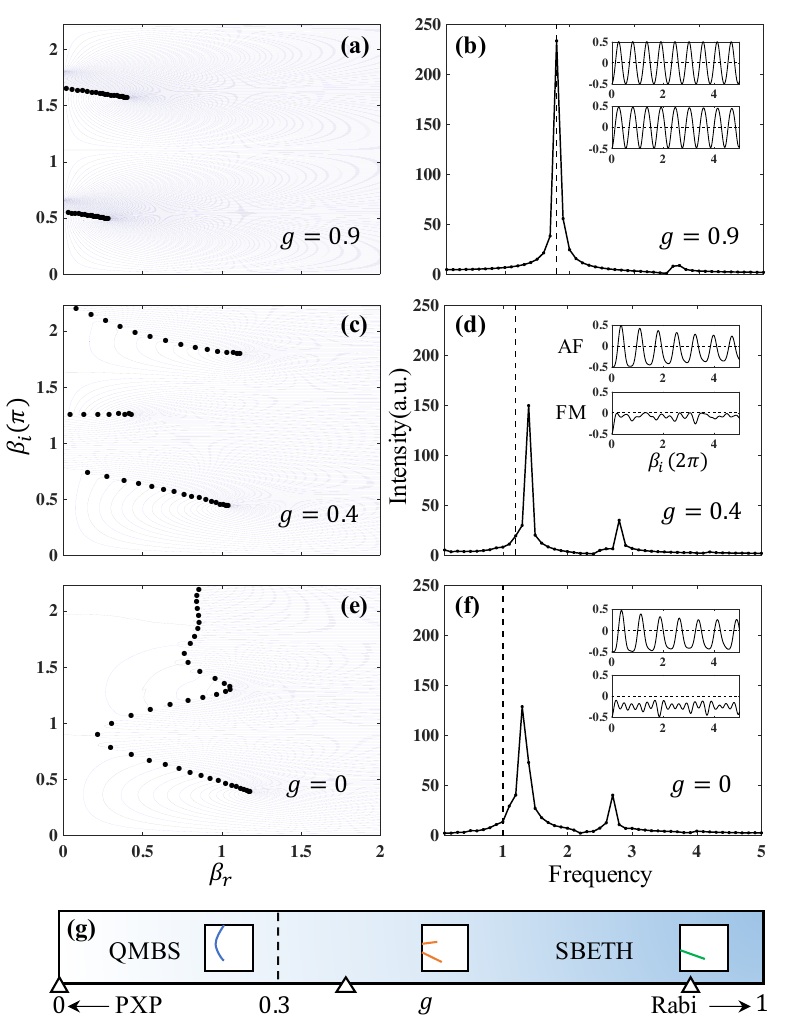}
    \caption{Fisher zeroes and real-time dynamics in the $\bar{P}X\bar{P}$ model:
(a), (c), and (e) depict the Fisher zeros computed from HOTRG for $g=0.9$, 0.4, and 0, respectively ($L=64$). The blue and gray lines represent the solutions of $\mathrm{Re}Z=0$ and $\mathrm{Im}Z=0$. Their intersections (black dots) give the Fisher zeroes. (b), (d), and (f) show the frequency spectra of the AFM state dynamics obtained from ED for system size $L=12$. The vertical dashed lines correspond to the first energy gap. The insets of each panel display the real-time oscillation curves of the AFM (upper) and FM states (lower), with time in units of $2\pi$. (g) The two distinct regions of thermalization behavior: QMBS ($0\le g\lesssim 0.3$) and SBETH ($0.3\lesssim g<1$).}
    \label{fig:fig2}
\end{figure}

We employ the tensor network method~\cite{ORUS2014117,TNreview1,TNreview2} to compute the complex $Z$ and its Fisher zeros. Specifically, using Trotter decomposition, the partition function of a one-dimensional quantum system of length $L$ can be mapped to a two-dimensional tensor network of size 
$L\times N$, where $N=\beta/\tau$ with the Trotter time $\tau$~\cite{Suzuki1976}. In our calculations, we fix $N=1024$.
We utilize the high order tensor renormalization group (HOTRG) algorithm~\cite{XieHOTRG} extended to complex $\beta$~\cite{Zou2014PRD} to contract this two-dimensional network and determine the locations where the real and imaginary parts of $Z$ vanish, thereby identifying the Fisher zeros through their intersections. The density of Fisher zeros increases with $L$ and usually form continuous lines in the thermodynamic limit~\cite{supp}. 
To capture nonthermal oscillatory dynamics, we use exact diagonalization (ED) to perform long-time dynamical evolution $|\psi(\beta_i)\rangle=\exp(-i\beta_i H)|\psi_0\rangle$ with a chosen initial state $\psi_0$, e.g., the antiferromagnetic (AFM) or ferromagnetic (FM) state. Here $\beta_i$ goes up to $40\pi$ and plays the role of time $t$, and the system size $L=12$. We measure $\langle S_z \rangle =\langle\psi(\beta_i)|S_z|\psi(\beta_i)\rangle$ with $S_z=\sigma_z/2$ at any given site. 
The numerical results are shown in Fig.~\ref{fig:fig2}.

At $g=1$, the Fisher zeros are located at isolated points $\beta_i=\pi/2+n\pi$ on the $\beta_i$-axis. Correspondingly, all states, including AFM and FM, undergo Rabi oscillations.
As $g$ is decreased, the interaction effects kick in and the Fisher zeros extend from the $\beta_i$ axis into the complex $\beta$ plane to form small line segments. The solutions of $\mathrm{Re}Z=0$ and $\mathrm{Im}Z=0$ remain dense near the $\beta_i$ axis [Fig.~\ref{fig:fig2}(a)]. Both AFM and FM states exhibit oscillations, with frequencies corresponding to the system's low-energy excitations.
In the small $\beta_r$ region near the $\beta_i$ axis, a vertical path will cut through these lines of zeros, suggesting that thermalization is hindered at long times. 
The evasion of thermalization is closely related to the long-lived low-energy excitations.
Therefore, this parameter regime can be identified as being dominated by SBETH [Fig.~\ref{fig:fig2}(b)].
As $g$ further decreases, additional line segments of Fisher zeros emerge, 
while the density of the solutions of $\mathrm{Re}Z=0$ and $\mathrm{Im}Z=0$ gradually decreases 
[Fig.~\ref{fig:fig2}(c)]. Accordingly, the main oscillation frequency of the AFM state shifts away from the low-energy excitations, and a secondary peak emerges at roughly twice the frequency, reminiscent of the tower structure of scars. In contrast, the FM state
ceases to oscillate with only weak spin-flip effects retained. This regime can be viewed as the crossover region between SBETH and QMBS [Fig.~\ref{fig:fig2}(d)]. Around $g\approx 0.3$, the Fisher zero lines begin to detach from the $\beta_i$ axis and connect vertically [Fig.~\ref{fig:fig2}(e)]. Dynamically, only the AFM state maintains nonthermal oscillations, so the parameter regime $g\lesssim 0.3$ is dominated by QMBS [Fig.~\ref{fig:fig2}(f)]. Figure.~\ref{fig:fig2}(g) summarizes the crossover from QMBS to SBETH as $g$ is varied, with each region depicted by its distinct structures of Fisher zeros.
More theoretical arguments for why Fisher zeros are a general diagnostic tool for QMBS are presented in the End Matter.

{~\it Ising chain with external fields—}To demonstrate the rich interplay between QMBS and SBETH and the utility of diagnosing QMBS through Fisher zeros, we examine another many-body Hamiltonian, the one-dimensional Ising chain in external fields with PBC,
\begin{equation}
    {H} = -\sum_{j=1}^{L} \left(\sigma_j^z\sigma_{j+1}^z + g_t\sigma_j^x + h_l\sigma_j^z\right),
    \label{eq:HamTLFIM}
\end{equation}
where $g_t$ and $h_l$ are the strength of the transverse and longitudinal fields, respectively.

\begin{figure}[t]
    \includegraphics[width=0.49\textwidth]{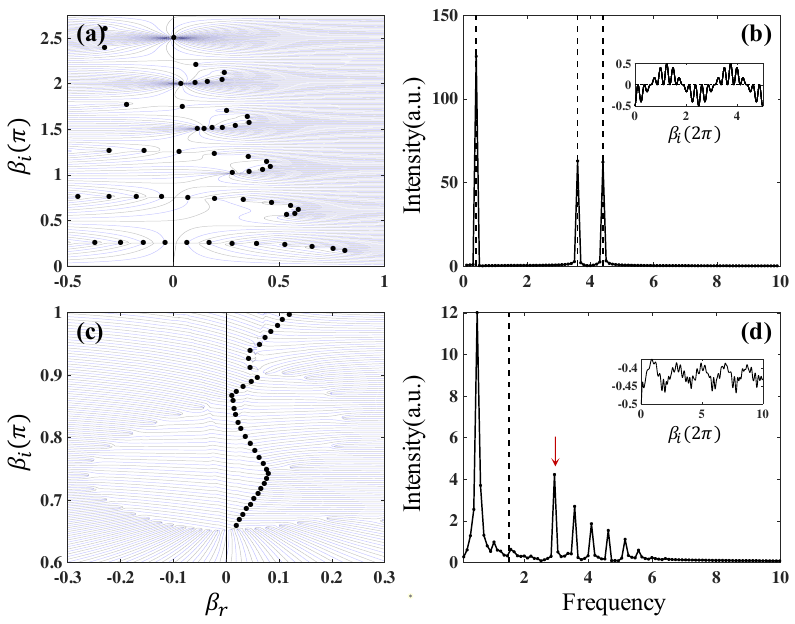}%
    \caption{Fisher zeros and real-time dynamics of the Ising chain Eq. \eqref{eq:HamTLFIM}. 
    (a) The Fisher zeros (black dots) for $g_t=0$, $h_l=0.2$ ($L=32$). (b) The frequency spectrum of the $|\cdots\rightarrow\cdots\rangle$ state for $L=12$. Dashed lines correspond to the energy levels in the spectrum. (c) The Fisher zeros of the case $g_t=0.5$, $h_l=0.1$, and $L=128$. Only the zeros describing the scars are marked by black dots. (d) The frequency spectrum of the FM state. Dashed lines correspond to the first excited state. The red arrow points to the scar peak, while other peaks arising from finite-size effects form a tower structure.}
    \label{fig:fig3}
\end{figure}

We first consider the  $g_t=0$ case with only the longitudinal field present. In this limit $Z$ and Fisher zeros can be determined exactly. When $h_l=0$, the Fisher zeros form lines parallel to the $\beta_r$ axis, extending to $\beta_r=+\infty$. Once $h_l$ is introduced, these zeros terminate at finite values of $\beta_r$, yielding a structure similar to that observed near $g=1$ in the $\bar{P}X\bar{P}$ model. Figure~\ref{fig:fig3}(a) shows the configuration of Fisher zeros (black dots) for the case $h_l=0.2$: in the complex $\beta$ plane, they cross the $\beta_i$ axis in periodic fashion. The termination points of the zeros from different line sectors form an envelope curve with a larger period, indicating a proportional relationship between the two periodicities. We select $|\cdots\rightarrow\cdots\rangle$, i.e., the direct product of all the $\sigma^x_j$ eigenstates, 
and illustrate its dynamical evolution (inset) and the corresponding frequency spectrum in Fig.~\ref{fig:fig3}(b). The intricate SBETH oscillations have frequencies matching certain excitations (indicated by dash lines) in the system's spectrum. These frequency modes also align with the periodicities of the Fisher zeros and the envelope curve in Fig.~\ref{fig:fig3}(a).

QMBS emerges when both $h_l$ and $g_t$ are present. The exact locations of Fisher zeros can be found for $h_l=0$~\cite{Liu2023CPL}. For instance, for $g_t<1$, the open lines and closed loops of zeros alternately appear in the $\beta$ plane. When $h_l$ is introduced, the closed loops break open, setting free the meson quasiparticles previously confined within the loops~\cite{Liu2023CPL}. 
This is shown in Fig.~\ref{fig:fig3}(c) for the example of $g_t=0.5$ and $h_l=0.1$.
Correspondingly, a set of Fisher zeros  are displaced away from the $\beta_i$ axis by $h_l$ to form a continuous line (black dots). This is the hallmark of QMBS, and rather similar to  the $\bar{P}X\bar{P}$ model near $g=0$.
The coexistence of these two types of Fisher zeros, namely, the broken loops and the emergent line extending
to large $\beta_i$, suggests that in strongly correlated models, quasiparticles can influence scar phenomena. 
We choose the FM state, which does not undergo DQPT, as an example of the time evolution. Its oscillatory behavior and the corresponding frequencies are shown in Fig.~\ref{fig:fig3}(d). We find that the peak at frequency 2.9, 
marked by a red arrow, remains unchanged with system size and away from the system's lowest-energy excitation. This peak corresponds exactly to the scar effect induced by meson excitations~\cite{Robinson2019PRL}. 
Although other nearly equally spaced peaks arise due to finite-size
effects, they build upon the meson peak~\cite{supp}.

{~\it Generalization and outlook—}We have confirmed our conjecture which relates Fisher zeros to QMBS
using two models, Eqs. \eqref{eq:HamPXP} and \eqref{eq:HamTLFIM}. We have also identified the distinct characteristics of weak and strong ergodicity breaking: for QMBS the Fisher zeros are off the $\beta_i$ axis extending to large $\beta_i$, while for SBETH the Fisher zeros intersect the $\beta_i$ axis. This approach
can be applied to other models. One example is provided in~\cite{supp}, where we solve 
a cluster spin model closely related to 1+1D lattice gauge theory~\cite{Sergej2020prl} that exhibits QMBS~\cite{Iadecola2020prb,Halimeh2023Quantum}. Its Fisher zero configurations again support the validity of our conjecture. These three model studies offer strong evidence that the unifying link between Fisher zeros and thermalization breakdown is rather general.

Our work suggests that Fisher zeros 
carry more theoretical significance than previously recognized.  
The knowledge of Fisher zeros can aid the construction of parent Hamiltonians~\cite{AKLT1987,Yang2023Construction,Zollar2024} to achieve novel quantum phases of matter. 
Nonunitary evolution in monitored quantum circuits~\cite{Francis2021SA,fisher2023random,Granet2023prl} offers a natural way for realization and detection of complex $Z$. In addition, the rapidly improving capacity to measure the many-body states of Rydberg atoms~\cite{Lukin2017Nature,Lukin2021Science} promises probing Fisher zeros in finite systems. 
For example, one can extract the effective temperature $1/\beta_r$ by fitting the decay of the Loschmidt ratio for different states. Summing over these states gives the total $Z$ to reveal at which time it reaches zero. Given that Lee-Yang zeros have been detected by tuning external fields~\cite{experimentLeeYang}, it is reasonable to expect the experimental observation of Fisher zeros in the near future.

{~\it Acknowledgments—}We thank Xiwen Guan, Gaoyong Sun, Junsen Wang, and Li You for helpful discussions. 
H.Z. is supported by the National Natural Science Foundation of China (Grant No. 12274126). 
E.Z. acknowledges the support from NSF Grant No. PHY-206419, and AFOSR Grant No. FA9550-23-1-0598. We conducted part of this work during the 2nd Conference on ``Quantum Simulations of Fundamental Physics" and the International Workshop on ``Quantum Systems with Novel Spatiotemporal Control" in Shanghai.

{~\it Data availability.} The data that support the findings of this article are openly available~\cite{figdata}.

\begin{widetext}
\section{End Matter}
\end{widetext}

\makeatletter

\begin{figure}[t]
    \includegraphics[width=0.49\textwidth]{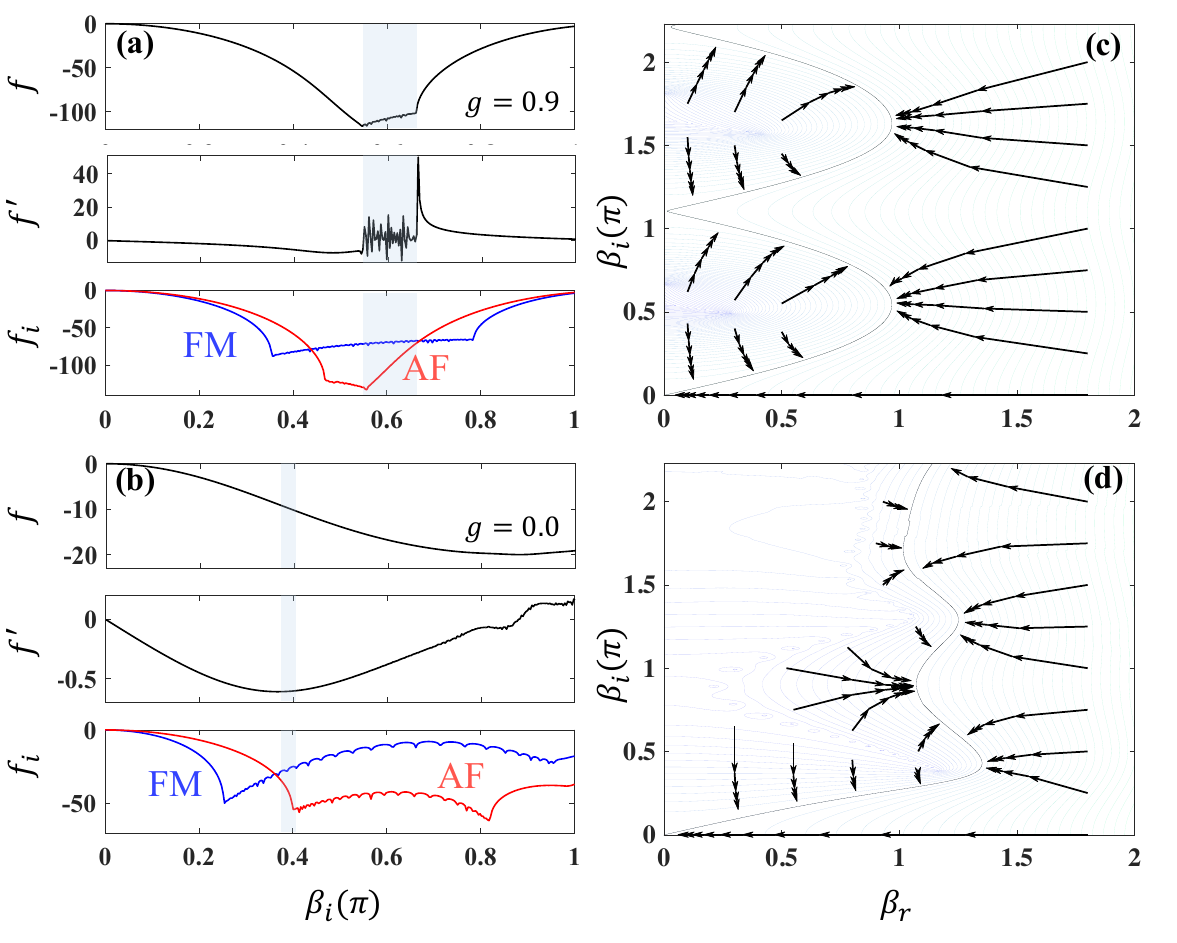}%
    \caption{Key differences between SBETH and QMBS in the analytical structure of $Z$ and the renormalization group (RG) flow. Left: $f=\ln |Z|$, $f'\equiv df/d\beta_i$, and $f_i=\ln |Z_i|$ for the AFM (red) and FM (blue) states as functions of $\beta_i$ with $\beta_r=0$. In the SBETH case (a, $g=0.9$), $f$ becomes nonanalytic over a broad interval of $\beta_i$. In the QMBS case (b, $g=0$), both $f$ and $f'$ evolve smoothly and $f$ exhibits only an
    inflection point. Right: the RG flows (arrows) for $g=0.9$ (c) and $g=0$ (d). The black lines connecting to the origin are the fixed points defined by $S_0=1$. The contour lines are for different values of $S_0$ at $L=64$.}
    \label{fig:fig4}
\end{figure}

{\bf Appendix A: Mountain range and water basin—}To gain a deeper understanding of the location of Fisher zeros for QMBS
and its qualitative difference from SBETH, we examine the analytical properties of $Z$ along the $\beta_i$ axis.
Recall $Z=\sum_i Z_i$, where $\{|\psi_i\rangle\}$ form a complete basis and the zeros of $Z_i$ (DQPT) mark the singularities of $f_i=\ln |Z_i|$.
In the case of SBETH [Fig.~\ref{fig:fig4}(a)], $f$ becomes nonanalytic over finite intervals (shaded regions), due to the collective effect of a broad set of states developing close-packed singularities in $f_i$. Interestingly, one of the singular points of the AFM state (red curve) is very close to the boundary of the nonanalytic region.
In contrast, for the QMBS case both $f$ and $f'$ remain smooth [Fig.~\ref{fig:fig4}(b)], different $Z_i$ exhibit different patterns of zeros in the complex $\beta$ plane~\cite{supp}, and the dynamics of $Z_i(\beta_r=0,\beta_i)$ is also different for
scar and nonscar states. In this case, the singularities of the non-scar FM state have little effect on $f$, while a singular point of the scar AFM state coincides with the inflection point of $f$ (shaded narrow region). 
The inflection point also corresponds to the projection of the lowest Fisher zero (see Fig. 2e)
onto the $\beta_i$ axis. This ``shadow effect" is analogous to how the rightmost Fisher zeros near the $\beta_r$ axis influence thermodynamic quantities such as specific heat~\cite{Liu2024CPL}.

The distinction between strong and weak ergodicity breaking can also be elucidated from 
renormalization group (RG). It has been noted that Fisher zeros form the boundary of RG flows
on the complex $\beta$-plane~\cite{RGflow2010,Zou2011PRD}.
As shown in ~\cite{supp}, the ``equipotential line" defined by $S_0=|Z(\beta_r,t)/Z(0,0)|^2=1$ converges to a continuous line in $\beta$ plane as the system size $L\rightarrow \infty$, providing a convenient ``high-temperature" (since it is connected to the origin $\beta=0$) fixed line for RG. 
The general trend of RG flows in Fig.~\ref{fig:fig4} can be appreciated by a geological analogy:
water flows toward the $S_0=1$ line (river in water basin) from near the Fisher zeros (mountain range).
For SBETH, the Fisher zeros cross the $\beta_i$ axis, allowing the RG flow to reach the $S_0=1$ line throughout the $\beta$ plane [Fig.~\ref{fig:fig4}(c)]. 
The periodic return along the time ($\beta_i$) axis clearly violates the ETH.
In the QMBS case, there exists a large plateau region between the zero line and the $\beta_i$ axis where the RG flows (not shown) are blocked by the ``mountain range" of Fisher zeros and cannot reach the $S_0=1$ fixed line [Fig.~\ref{fig:fig4}(d)]. Thus the ``mountain range" of Fisher zeros divides the $\beta$ plane into regions with distinctive long-time behaviors in $Z$, in agreement with the intuitive argument used earlier to motivate our conjecture.

\begin{figure}[htbp]
    \includegraphics[width=0.49\textwidth]{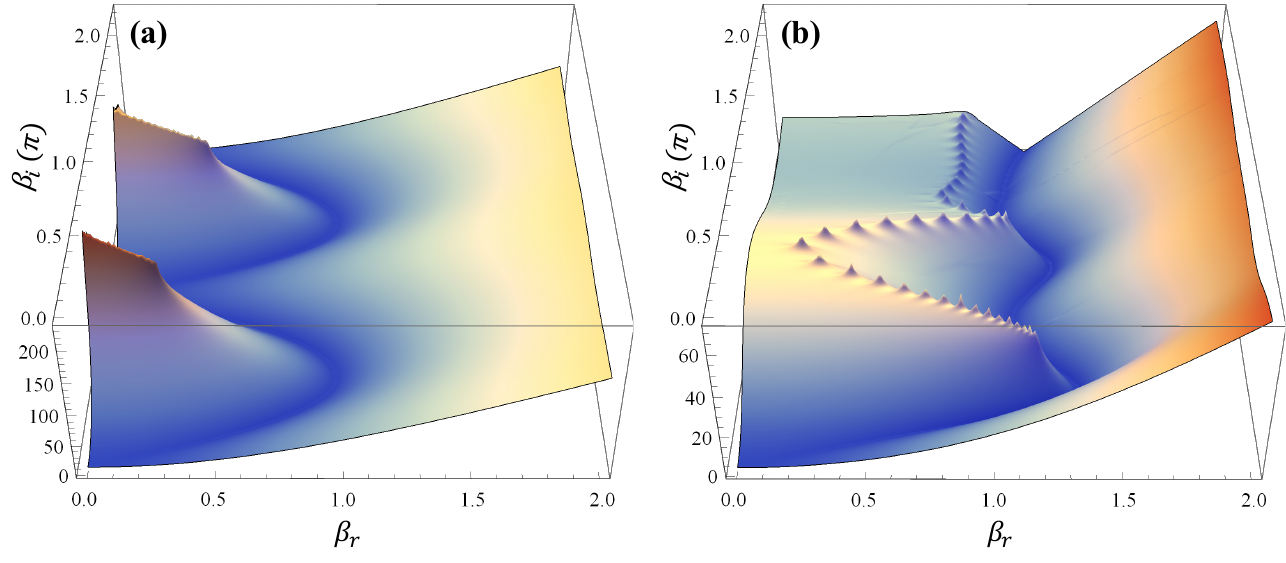}%
    \caption{Three-dimensional contour plots of $|\ln S_0|$ for $g=0.9$ (a) and $g=0$ (b).}
    \label{fig:fig5}
\end{figure}

We note that to the left of the $S_0=1$ line, $S_0<1$, and $S_0=0$ is reached precisely at the Fisher zero, while to the right of the $S_0=1$ line, $S_0>1$. The value of $|\ln S_0|$ reaches a minimum exactly at $S_0=1$ line, making this line the lowest point in the $|\ln S_0|$ landscape. Therefore, to more intuitively illustrate how the RG flow originates from either the Fisher zeros or large $\beta$ regions and approaches the $S_0=1$ line, we show in Fig.~\ref{fig:fig5} the values of $|\ln S_0|$ for $L=64$, which clearly captures the ``mountain range and water basinlike" structure as previously discussed.


\begin{thebibliography}{67}%
\makeatletter
\providecommand \@ifxundefined [1]{%
 \@ifx{#1\undefined}
}%
\providecommand \@ifnum [1]{%
 \ifnum #1\expandafter \@firstoftwo
 \else \expandafter \@secondoftwo
 \fi
}%
\providecommand \@ifx [1]{%
 \ifx #1\expandafter \@firstoftwo
 \else \expandafter \@secondoftwo
 \fi
}%
\providecommand \natexlab [1]{#1}%
\providecommand \enquote  [1]{``#1''}%
\providecommand \bibnamefont  [1]{#1}%
\providecommand \bibfnamefont [1]{#1}%
\providecommand \citenamefont [1]{#1}%
\providecommand \href@noop [0]{\@secondoftwo}%
\providecommand \href [0]{\begingroup \@sanitize@url \@href}%
\providecommand \@href[1]{\@@startlink{#1}\@@href}%
\providecommand \@@href[1]{\endgroup#1\@@endlink}%
\providecommand \@sanitize@url [0]{\catcode `\\12\catcode `\$12\catcode
  `\&12\catcode `\#12\catcode `\^12\catcode `\_12\catcode `\%12\relax}%
\providecommand \@@startlink[1]{}%
\providecommand \@@endlink[0]{}%
\providecommand \url  [0]{\begingroup\@sanitize@url \@url }%
\providecommand \@url [1]{\endgroup\@href {#1}{\urlprefix }}%
\providecommand \urlprefix  [0]{URL }%
\providecommand \Eprint [0]{\href }%
\providecommand \doibase [0]{https://doi.org/}%
\providecommand \selectlanguage [0]{\@gobble}%
\providecommand \bibinfo  [0]{\@secondoftwo}%
\providecommand \bibfield  [0]{\@secondoftwo}%
\providecommand \translation [1]{[#1]}%
\providecommand \BibitemOpen [0]{}%
\providecommand \bibitemStop [0]{}%
\providecommand \bibitemNoStop [0]{.\EOS\space}%
\providecommand \EOS [0]{\spacefactor3000\relax}%
\providecommand \BibitemShut  [1]{\csname bibitem#1\endcsname}%
\let\auto@bib@innerbib\@empty
\bibitem [{\citenamefont {Deutsch}(1991)}]{Deutsch1991PRA}%
  \BibitemOpen
  \bibfield  {author} {\bibinfo {author} {\bibfnamefont {J.~M.}\ \bibnamefont
  {Deutsch}},\ }\bibfield  {title} {\bibinfo {title} {Quantum statistical
  mechanics in a closed system},\ }\href
  {https://doi.org/10.1103/PhysRevA.43.2046} {\bibfield  {journal} {\bibinfo
  {journal} {Phys. Rev. A}\ }\textbf {\bibinfo {volume} {43}},\ \bibinfo
  {pages} {2046} (\bibinfo {year} {1991})}\BibitemShut {NoStop}%
\bibitem [{\citenamefont {Srednicki}(1994)}]{Srednicki1994PRE}%
  \BibitemOpen
  \bibfield  {author} {\bibinfo {author} {\bibfnamefont {M.}~\bibnamefont
  {Srednicki}},\ }\bibfield  {title} {\bibinfo {title} {Chaos and quantum
  thermalization},\ }\href {https://doi.org/10.1103/PhysRevE.50.888} {\bibfield
   {journal} {\bibinfo  {journal} {Phys. Rev. E}\ }\textbf {\bibinfo {volume}
  {50}},\ \bibinfo {pages} {888} (\bibinfo {year} {1994})}\BibitemShut
  {NoStop}%
\bibitem [{\citenamefont {Turner}\ \emph
  {et~al.}(2018{\natexlab{a}})\citenamefont {Turner}, \citenamefont
  {Michailidis}, \citenamefont {Abanin}, \citenamefont {Serbyn},\ and\
  \citenamefont {Papi\'{c}}}]{Papic2018NP}%
  \BibitemOpen
  \bibfield  {author} {\bibinfo {author} {\bibfnamefont {C.~J.}\ \bibnamefont
  {Turner}}, \bibinfo {author} {\bibfnamefont {A.~A.}\ \bibnamefont
  {Michailidis}}, \bibinfo {author} {\bibfnamefont {D.~A.}\ \bibnamefont
  {Abanin}}, \bibinfo {author} {\bibfnamefont {M.}~\bibnamefont {Serbyn}},\
  and\ \bibinfo {author} {\bibfnamefont {Z.}~\bibnamefont {Papi\'{c}}},\
  }\bibfield  {title} {\bibinfo {title} {Weak ergodicity breaking from quantum
  many-body scars},\ }\href {https://doi.org/10.1038/s41567-018-0137-5}
  {\bibfield  {journal} {\bibinfo  {journal} {Nat. Phys.}\ }\textbf
  {\bibinfo {volume} {14}},\ \bibinfo {pages} {745} (\bibinfo {year}
  {2018}{\natexlab{a}})}\BibitemShut {NoStop}%
\bibitem [{\citenamefont {Moudgalya}\ \emph {et~al.}(2018)\citenamefont
  {Moudgalya}, \citenamefont {Regnault},\ and\ \citenamefont
  {Bernevig}}]{Bernevig2018PRB}%
  \BibitemOpen
  \bibfield  {author} {\bibinfo {author} {\bibfnamefont {S.}~\bibnamefont
  {Moudgalya}}, \bibinfo {author} {\bibfnamefont {N.}~\bibnamefont
  {Regnault}},\ and\ \bibinfo {author} {\bibfnamefont {B.~A.}\ \bibnamefont
  {Bernevig}},\ }\bibfield  {title} {\bibinfo {title} {Entanglement of exact
  excited states of Affleck-Kennedy-Lieb-Tasaki models: Exact results,
  many-body scars, and violation of the strong eigenstate thermalization
  hypothesis},\ }\href {https://doi.org/10.1103/PhysRevB.98.235156} {\bibfield
  {journal} {\bibinfo  {journal} {Phys. Rev. B}\ }\textbf {\bibinfo {volume}
  {98}},\ \bibinfo {pages} {235156} (\bibinfo {year} {2018})}\BibitemShut
  {NoStop}%
\bibitem [{\citenamefont {Schecter}\ and\ \citenamefont
  {Iadecola}(2019)}]{Schecter2019PRL}%
  \BibitemOpen
  \bibfield  {author} {\bibinfo {author} {\bibfnamefont {M.}~\bibnamefont
  {Schecter}}\ and\ \bibinfo {author} {\bibfnamefont {T.}~\bibnamefont
  {Iadecola}},\ }\bibfield  {title} {\bibinfo {title} {Weak ergodicity breaking
  and quantum many-body scars in spin-1 $XY$ magnets},\ }\href
  {https://doi.org/10.1103/PhysRevLett.123.147201} {\bibfield  {journal}
  {\bibinfo  {journal} {Phys. Rev. Lett.}\ }\textbf {\bibinfo {volume} {123}},\
  \bibinfo {pages} {147201} (\bibinfo {year} {2019})}\BibitemShut {NoStop}%
\bibitem [{\citenamefont {James}\ \emph {et~al.}(2019)\citenamefont {James},
  \citenamefont {Konik},\ and\ \citenamefont {Robinson}}]{Robinson2019PRL}%
  \BibitemOpen
  \bibfield  {author} {\bibinfo {author} {\bibfnamefont {A.~J.~A.}\
  \bibnamefont {James}}, \bibinfo {author} {\bibfnamefont {R.~M.}\ \bibnamefont
  {Konik}},\ and\ \bibinfo {author} {\bibfnamefont {N.~J.}\ \bibnamefont
  {Robinson}},\ }\bibfield  {title} {\bibinfo {title} {Nonthermal states
  arising from confinement in one and two dimensions},\ }\href
  {https://doi.org/10.1103/PhysRevLett.122.130603} {\bibfield  {journal}
  {\bibinfo  {journal} {Phys. Rev. Lett.}\ }\textbf {\bibinfo {volume} {122}},\
  \bibinfo {pages} {130603} (\bibinfo {year} {2019})}\BibitemShut {NoStop}%
\bibitem [{\citenamefont {Sala}\ \emph {et~al.}(2020)\citenamefont {Sala},
  \citenamefont {Rakovszky}, \citenamefont {Verresen}, \citenamefont {Knap},\
  and\ \citenamefont {Pollmann}}]{Pollmann2020PRX}%
  \BibitemOpen
  \bibfield  {author} {\bibinfo {author} {\bibfnamefont {P.}~\bibnamefont
  {Sala}}, \bibinfo {author} {\bibfnamefont {T.}~\bibnamefont {Rakovszky}},
  \bibinfo {author} {\bibfnamefont {R.}~\bibnamefont {Verresen}}, \bibinfo
  {author} {\bibfnamefont {M.}~\bibnamefont {Knap}},\ and\ \bibinfo {author}
  {\bibfnamefont {F.}~\bibnamefont {Pollmann}},\ }\bibfield  {title} {\bibinfo
  {title} {Ergodicity breaking arising from Hilbert space fragmentation in
  dipole-conserving Hamiltonians},\ }\href
  {https://doi.org/10.1103/PhysRevX.10.011047} {\bibfield  {journal} {\bibinfo
  {journal} {Phys. Rev. X}\ }\textbf {\bibinfo {volume} {10}},\ \bibinfo
  {pages} {011047} (\bibinfo {year} {2020})}\BibitemShut {NoStop}%
\bibitem [{\citenamefont {Bernien}\ \emph {et~al.}(2017)\citenamefont
  {Bernien}, \citenamefont {Schwartz}, \citenamefont {Keesling}, \citenamefont
  {Levine}, \citenamefont {Omran}, \citenamefont {Pichler}, \citenamefont
  {Choi}, \citenamefont {Zibrov}, \citenamefont {Endres}, \citenamefont
  {Greiner}, \citenamefont {Vuleti\'{c}},\ and\ \citenamefont
  {Lukin}}]{Lukin2017Nature}%
  \BibitemOpen
  \bibfield  {author} {\bibinfo {author} {\bibfnamefont {H.}~\bibnamefont
  {Bernien}}, \bibinfo {author} {\bibfnamefont {S.}~\bibnamefont {Schwartz}},
  \bibinfo {author} {\bibfnamefont {A.}~\bibnamefont {Keesling}}, \bibinfo
  {author} {\bibfnamefont {H.}~\bibnamefont {Levine}}, \bibinfo {author}
  {\bibfnamefont {A.}~\bibnamefont {Omran}}, \bibinfo {author} {\bibfnamefont
  {H.}~\bibnamefont {Pichler}}, \bibinfo {author} {\bibfnamefont
  {S.}~\bibnamefont {Choi}}, \bibinfo {author} {\bibfnamefont {A.~S.}\
  \bibnamefont {Zibrov}}, \bibinfo {author} {\bibfnamefont {M.}~\bibnamefont
  {Endres}}, \bibinfo {author} {\bibfnamefont {M.}~\bibnamefont {Greiner}},
  \bibinfo {author} {\bibfnamefont {V.}~\bibnamefont {Vuleti\'{c}}},\ and\
  \bibinfo {author} {\bibfnamefont {M.~D.}\ \bibnamefont {Lukin}},\ }\bibfield
  {title} {\bibinfo {title} {Probing many-body dynamics on a 51-atom quantum
  simulator},\ }\href {https://doi.org/10.1038/nature24622} {\bibfield
  {journal} {\bibinfo  {journal} {Nature (London)}\ }\textbf {\bibinfo {volume} {551}},\
  \bibinfo {pages} {579} (\bibinfo {year} {2017})}\BibitemShut {NoStop}%
\bibitem [{\citenamefont {Bluvstein}\ \emph {et~al.}(2021)\citenamefont
  {Bluvstein}, \citenamefont {Omran}, \citenamefont {Levine}, \citenamefont
  {Keesling}, \citenamefont {Semeghini}, \citenamefont {Ebadi}, \citenamefont
  {Wang}, \citenamefont {Michailidis}, \citenamefont {Maskara}, \citenamefont
  {Ho}, \citenamefont {Choi}, \citenamefont {Serbyn}, \citenamefont {Greiner},
  \citenamefont {Vuleti\'{c}},\ and\ \citenamefont {Lukin}}]{Lukin2021Science}%
  \BibitemOpen
  \bibfield  {author} {\bibinfo {author} {\bibfnamefont {D.}~\bibnamefont
  {Bluvstein}}, \bibinfo {author} {\bibfnamefont {A.}~\bibnamefont {Omran}},
  \bibinfo {author} {\bibfnamefont {H.}~\bibnamefont {Levine}}, \bibinfo
  {author} {\bibfnamefont {A.}~\bibnamefont {Keesling}}, \bibinfo {author}
  {\bibfnamefont {G.}~\bibnamefont {Semeghini}}, \bibinfo {author}
  {\bibfnamefont {S.}~\bibnamefont {Ebadi}}, \bibinfo {author} {\bibfnamefont
  {T.~T.}\ \bibnamefont {Wang}}, \bibinfo {author} {\bibfnamefont {A.~A.}\
  \bibnamefont {Michailidis}}, \bibinfo {author} {\bibfnamefont
  {N.}~\bibnamefont {Maskara}}, \bibinfo {author} {\bibfnamefont {W.~W.}\
  \bibnamefont {Ho}}, \bibinfo {author} {\bibfnamefont {S.}~\bibnamefont
  {Choi}}, \bibinfo {author} {\bibfnamefont {M.}~\bibnamefont {Serbyn}},
  \bibinfo {author} {\bibfnamefont {M.}~\bibnamefont {Greiner}}, \bibinfo
  {author} {\bibfnamefont {V.}~\bibnamefont {Vuleti\'{c}}},\ and\ \bibinfo
  {author} {\bibfnamefont {M.~D.}\ \bibnamefont {Lukin}},\ }\bibfield  {title}
  {\bibinfo {title} {Controlling quantum many-body dynamics in driven Rydberg
  atom arrays},\ }\href {https://doi.org/10.1126/science.abg2530} {\bibfield
  {journal} {\bibinfo  {journal} {Science}\ }\textbf {\bibinfo {volume}
  {371}},\ \bibinfo {pages} {1355} (\bibinfo {year} {2021})}\BibitemShut
  {NoStop}%
\bibitem [{\citenamefont {Zhang}\ \emph {et~al.}(2022)\citenamefont {Zhang},
  \citenamefont {Dong}, \citenamefont {Gao}, \citenamefont {Zhao},
  \citenamefont {Hao}, \citenamefont {Desaules}, \citenamefont {Guo},
  \citenamefont {Chen}, \citenamefont {Deng}, \citenamefont {Liu},
  \citenamefont {Ren}, \citenamefont {Yao}, \citenamefont {Zhang},
  \citenamefont {Xu}, \citenamefont {Wang}, \citenamefont {Jin}, \citenamefont
  {Zhu}, \citenamefont {Zhang}, \citenamefont {Li}, \citenamefont {Song},
  \citenamefont {Wang}, \citenamefont {Liu}, \citenamefont {Papi\'{c}},
  \citenamefont {Ying}, \citenamefont {Wang},\ and\ \citenamefont
  {Lai}}]{Zhang2022NP}%
  \BibitemOpen
  \bibfield  {author} {\bibinfo {author} {\bibfnamefont {P.}~\bibnamefont
  {Zhang}}, \bibinfo {author} {\bibfnamefont {H.}~\bibnamefont {Dong}},
  \bibinfo {author} {\bibfnamefont {Y.}~\bibnamefont {Gao}}, \bibinfo {author}
  {\bibfnamefont {L.}~\bibnamefont {Zhao}}, \bibinfo {author} {\bibfnamefont
  {J.}~\bibnamefont {Hao}}, \bibinfo {author} {\bibfnamefont {J.-Y.}\
  \bibnamefont {Desaules}}, \bibinfo {author} {\bibfnamefont {Q.}~\bibnamefont
  {Guo}}, \bibinfo {author} {\bibfnamefont {J.}~\bibnamefont {Chen}}, \bibinfo
  {author} {\bibfnamefont {J.}~\bibnamefont {Deng}}, \bibinfo {author}
  {\bibfnamefont {B.}~\bibnamefont {Liu}}, \bibinfo {author} {\bibfnamefont
  {W.}~\bibnamefont {Ren}}, \bibinfo {author} {\bibfnamefont {Y.}~\bibnamefont
  {Yao}}, \bibinfo {author} {\bibfnamefont {X.}~\bibnamefont {Zhang}}, \bibinfo
  {author} {\bibfnamefont {S.}~\bibnamefont {Xu}}, \bibinfo {author}
  {\bibfnamefont {K.}~\bibnamefont {Wang}}, \bibinfo {author} {\bibfnamefont
  {F.}~\bibnamefont {Jin}}, \bibinfo {author} {\bibfnamefont {X.}~\bibnamefont
  {Zhu}}, \bibinfo {author} {\bibfnamefont {B.}~\bibnamefont {Zhang}}, \bibinfo
  {author} {\bibfnamefont {H.}~\bibnamefont {Li}}, \bibinfo {author}
  {\bibfnamefont {C.}~\bibnamefont {Song}}, \bibinfo {author} {\bibfnamefont
  {Z.}~\bibnamefont {Wang}}, \bibinfo {author} {\bibfnamefont {F.}~\bibnamefont
  {Liu}}, \bibinfo {author} {\bibfnamefont {Z.}~\bibnamefont {Papi\'{c}}},
  \bibinfo {author} {\bibfnamefont {L.}~\bibnamefont {Ying}}, \bibinfo {author}
  {\bibfnamefont {H.}~\bibnamefont {Wang}},\ and\ \bibinfo {author}
  {\bibfnamefont {Y.-C.}\ \bibnamefont {Lai}},\ }\bibfield  {title} {\bibinfo
  {title} {Many-body Hilbert space scarring on a superconducting processor},\
  }\href {https://doi.org/10.1038/s41567-022-01784-9} {\bibfield  {journal}
  {\bibinfo  {journal} {Nat. Phys.}\ }\textbf {\bibinfo {volume} {19}},\
  \bibinfo {pages} {120} (\bibinfo {year} {2022})}\BibitemShut {NoStop}%
\bibitem [{\citenamefont {Su}\ \emph {et~al.}(2023)\citenamefont {Su},
  \citenamefont {Sun}, \citenamefont {Hudomal}, \citenamefont {Desaules},
  \citenamefont {Zhou}, \citenamefont {Yang}, \citenamefont {Halimeh},
  \citenamefont {Yuan}, \citenamefont {Papi\'{c}},\ and\ \citenamefont
  {Pan}}]{Pan2023PRR_QMBS}%
  \BibitemOpen
  \bibfield  {author} {\bibinfo {author} {\bibfnamefont {G.-X.}\ \bibnamefont
  {Su}}, \bibinfo {author} {\bibfnamefont {H.}~\bibnamefont {Sun}}, \bibinfo
  {author} {\bibfnamefont {A.}~\bibnamefont {Hudomal}}, \bibinfo {author}
  {\bibfnamefont {J.-Y.}\ \bibnamefont {Desaules}}, \bibinfo {author}
  {\bibfnamefont {Z.-Y.}\ \bibnamefont {Zhou}}, \bibinfo {author}
  {\bibfnamefont {B.}~\bibnamefont {Yang}}, \bibinfo {author} {\bibfnamefont
  {J.~C.}\ \bibnamefont {Halimeh}}, \bibinfo {author} {\bibfnamefont {Z.-S.}\
  \bibnamefont {Yuan}}, \bibinfo {author} {\bibfnamefont {Z.}~\bibnamefont
  {Papi\'{c}}},\ and\ \bibinfo {author} {\bibfnamefont {J.-W.}\ \bibnamefont
  {Pan}},\ }\bibfield  {title} {\bibinfo {title} {Observation of many-body
  scarring in a Bose-Hubbard quantum simulator},\ }\href
  {https://doi.org/10.1103/PhysRevResearch.5.023010} {\bibfield  {journal}
  {\bibinfo  {journal} {Phys. Rev. Res.}\ }\textbf {\bibinfo {volume} {5}},\
  \bibinfo {pages} {023010} (\bibinfo {year} {2023})}\BibitemShut {NoStop}%
\bibitem [{\citenamefont {Serbyn}\ \emph {et~al.}(2021)\citenamefont {Serbyn},
  \citenamefont {Abanin},\ and\ \citenamefont {Papić}}]{QMBSreview2021NP}%
  \BibitemOpen
  \bibfield  {author} {\bibinfo {author} {\bibfnamefont {M.}~\bibnamefont
  {Serbyn}}, \bibinfo {author} {\bibfnamefont {D.~A.}\ \bibnamefont {Abanin}},\
  and\ \bibinfo {author} {\bibfnamefont {Z.}~\bibnamefont {Papić}},\
  }\bibfield  {title} {\bibinfo {title} {Quantum many-body scars and weak
  breaking of ergodicity},\ }\href {https://doi.org/10.1038/s41567-021-01230-2}
  {\bibfield  {journal} {\bibinfo  {journal} {Nat. Phys.}\ }\textbf
  {\bibinfo {volume} {17}},\ \bibinfo {pages} {675} (\bibinfo {year}
  {2021})}\BibitemShut {NoStop}%
\bibitem [{\citenamefont {Moudgalya}\ \emph {et~al.}(2022)\citenamefont
  {Moudgalya}, \citenamefont {Bernevig},\ and\ \citenamefont
  {Regnault}}]{Moudgalya2022QMBSreview}%
  \BibitemOpen
  \bibfield  {author} {\bibinfo {author} {\bibfnamefont {S.}~\bibnamefont
  {Moudgalya}}, \bibinfo {author} {\bibfnamefont {B.~A.}\ \bibnamefont
  {Bernevig}},\ and\ \bibinfo {author} {\bibfnamefont {N.}~\bibnamefont
  {Regnault}},\ }\bibfield  {title} {\bibinfo {title} {Quantum many-body scars
  and Hilbert space fragmentation: A review of exact results},\ }\href
  {https://doi.org/10.1088/1361-6633/ac73a0} {\bibfield  {journal} {\bibinfo
  {journal} {Rep. Prog. Phys.}\ }\textbf {\bibinfo {volume}
  {85}},\ \bibinfo {pages} {086501} (\bibinfo {year} {2022})}\BibitemShut
  {NoStop}%
\bibitem [{\citenamefont {Moudgalya}\ and\ \citenamefont
  {Motrunich}(2024)}]{Moudgalya2024PRX}%
  \BibitemOpen
  \bibfield  {author} {\bibinfo {author} {\bibfnamefont {S.}~\bibnamefont
  {Moudgalya}}\ and\ \bibinfo {author} {\bibfnamefont {O.~I.}\ \bibnamefont
  {Motrunich}},\ }\bibfield  {title} {\bibinfo {title} {Exhaustive
  characterization of quantum many-body scars using commutant algebras},\
  }\href {https://doi.org/10.1103/PhysRevX.14.041069} {\bibfield  {journal}
  {\bibinfo  {journal} {Phys. Rev. X}\ }\textbf {\bibinfo {volume} {14}},\
  \bibinfo {pages} {041069} (\bibinfo {year} {2024})}\BibitemShut {NoStop}%
\bibitem [{\citenamefont {Shiraishi}\ and\ \citenamefont
  {Mori}(2017)}]{Shiraishi2017Embedding}%
  \BibitemOpen
  \bibfield  {author} {\bibinfo {author} {\bibfnamefont {N.}~\bibnamefont
  {Shiraishi}}\ and\ \bibinfo {author} {\bibfnamefont {T.}~\bibnamefont
  {Mori}},\ }\bibfield  {title} {\bibinfo {title} {Systematic construction of
  counterexamples to the eigenstate thermalization hypothesis},\ }\href
  {https://doi.org/10.1103/PhysRevLett.119.030601} {\bibfield  {journal}
  {\bibinfo  {journal} {Phys. Rev. Lett.}\ }\textbf {\bibinfo {volume} {119}},\
  \bibinfo {pages} {030601} (\bibinfo {year} {2017})}\BibitemShut {NoStop}%
\bibitem [{\citenamefont {Chandran}\ \emph {et~al.}(2023)\citenamefont
  {Chandran}, \citenamefont {Iadecola}, \citenamefont {Khemani},\ and\
  \citenamefont {Moessner}}]{QMBS2023ARCMP}%
  \BibitemOpen
  \bibfield  {author} {\bibinfo {author} {\bibfnamefont {A.}~\bibnamefont
  {Chandran}}, \bibinfo {author} {\bibfnamefont {T.}~\bibnamefont {Iadecola}},
  \bibinfo {author} {\bibfnamefont {V.}~\bibnamefont {Khemani}},\ and\ \bibinfo
  {author} {\bibfnamefont {R.}~\bibnamefont {Moessner}},\ }\bibfield  {title}
  {\bibinfo {title} {Quantum many-body scars: A quasiparticle perspective},\
  }\href {https://doi.org/10.1146/annurev-conmatphys-031620-101617} {\bibfield
  {journal} {\bibinfo  {journal} {Annu. Rev. Condens. Matter Phys.}\
  }\textbf {\bibinfo {volume} {14}},\ \bibinfo {pages} {443} (\bibinfo
  {year} {2023})}\BibitemShut {NoStop}%
\bibitem [{\citenamefont {Ba\~nuls}\ \emph {et~al.}(2011)\citenamefont
  {Ba\~nuls}, \citenamefont {Cirac},\ and\ \citenamefont
  {Hastings}}]{Banuls2011PRL}%
  \BibitemOpen
  \bibfield  {author} {\bibinfo {author} {\bibfnamefont {M.~C.}\ \bibnamefont
  {Ba\~nuls}}, \bibinfo {author} {\bibfnamefont {J.~I.}\ \bibnamefont
  {Cirac}},\ and\ \bibinfo {author} {\bibfnamefont {M.~B.}\ \bibnamefont
  {Hastings}},\ }\bibfield  {title} {\bibinfo {title} {Strong and weak
  thermalization of infinite nonintegrable quantum systems},\ }\href
  {https://doi.org/10.1103/PhysRevLett.106.050405} {\bibfield  {journal}
  {\bibinfo  {journal} {Phys. Rev. Lett.}\ }\textbf {\bibinfo {volume} {106}},\
  \bibinfo {pages} {050405} (\bibinfo {year} {2011})}\BibitemShut {NoStop}%
\bibitem [{\citenamefont {Kormos}\ \emph {et~al.}(2016)\citenamefont {Kormos},
  \citenamefont {Collura}, \citenamefont {Takács},\ and\ \citenamefont
  {Calabrese}}]{Kormos2016NP}%
  \BibitemOpen
  \bibfield  {author} {\bibinfo {author} {\bibfnamefont {M.}~\bibnamefont
  {Kormos}}, \bibinfo {author} {\bibfnamefont {M.}~\bibnamefont {Collura}},
  \bibinfo {author} {\bibfnamefont {G.}~\bibnamefont {Takács}},\ and\ \bibinfo
  {author} {\bibfnamefont {P.}~\bibnamefont {Calabrese}},\ }\bibfield  {title}
  {\bibinfo {title} {Real-time confinement following a quantum quench to a
  non-integrable model},\ }\href {https://doi.org/10.1038/nphys3934} {\bibfield
   {journal} {\bibinfo  {journal} {Nat. Phys.}\ }\textbf {\bibinfo {volume}
  {13}},\ \bibinfo {pages} {246} (\bibinfo {year} {2016})}\BibitemShut
  {NoStop}%
\bibitem [{\citenamefont {Lin}\ and\ \citenamefont
  {Motrunich}(2017)}]{Motrunich2017PRA}%
  \BibitemOpen
  \bibfield  {author} {\bibinfo {author} {\bibfnamefont {C.-J.}\ \bibnamefont
  {Lin}}\ and\ \bibinfo {author} {\bibfnamefont {O.~I.}\ \bibnamefont
  {Motrunich}},\ }\bibfield  {title} {\bibinfo {title} {Quasiparticle
  explanation of the weak-thermalization regime under quench in a nonintegrable
  quantum spin chain},\ }\href {https://doi.org/10.1103/PhysRevA.95.023621}
  {\bibfield  {journal} {\bibinfo  {journal} {Phys. Rev. A}\ }\textbf {\bibinfo
  {volume} {95}},\ \bibinfo {pages} {023621} (\bibinfo {year}
  {2017})}\BibitemShut {NoStop}%
\bibitem [{\citenamefont {Collura}\ \emph {et~al.}(2018)\citenamefont
  {Collura}, \citenamefont {Kormos},\ and\ \citenamefont
  {Tak\'acs}}]{Collura2018PRA}%
  \BibitemOpen
  \bibfield  {author} {\bibinfo {author} {\bibfnamefont {M.}~\bibnamefont
  {Collura}}, \bibinfo {author} {\bibfnamefont {M.}~\bibnamefont {Kormos}},\
  and\ \bibinfo {author} {\bibfnamefont {G.}~\bibnamefont {Tak\'acs}},\
  }\bibfield  {title} {\bibinfo {title} {Dynamical manifestation of the Gibbs
  paradox after a quantum quench},\ }\href
  {https://doi.org/10.1103/PhysRevA.98.053610} {\bibfield  {journal} {\bibinfo
  {journal} {Phys. Rev. A}\ }\textbf {\bibinfo {volume} {98}},\ \bibinfo
  {pages} {053610} (\bibinfo {year} {2018})}\BibitemShut {NoStop}%
\bibitem [{\citenamefont {Wilczek}(2012)}]{Wilczek2012QTC}%
  \BibitemOpen
  \bibfield  {author} {\bibinfo {author} {\bibfnamefont {F.}~\bibnamefont
  {Wilczek}},\ }\bibfield  {title} {\bibinfo {title} {Quantum time crystals},\
  }\href {https://doi.org/10.1103/PhysRevLett.109.160401} {\bibfield  {journal}
  {\bibinfo  {journal} {Phys. Rev. Lett.}\ }\textbf {\bibinfo {volume} {109}},\
  \bibinfo {pages} {160401} (\bibinfo {year} {2012})}\BibitemShut {NoStop}%
\bibitem [{\citenamefont {Else}\ \emph {et~al.}(2017)\citenamefont {Else},
  \citenamefont {Bauer},\ and\ \citenamefont {Nayak}}]{Nayak2017PRX}%
  \BibitemOpen
  \bibfield  {author} {\bibinfo {author} {\bibfnamefont {D.~V.}\ \bibnamefont
  {Else}}, \bibinfo {author} {\bibfnamefont {B.}~\bibnamefont {Bauer}},\ and\
  \bibinfo {author} {\bibfnamefont {C.}~\bibnamefont {Nayak}},\ }\bibfield
  {title} {\bibinfo {title} {Prethermal phases of matter protected by
  time-translation symmetry},\ }\href
  {https://doi.org/10.1103/PhysRevX.7.011026} {\bibfield  {journal} {\bibinfo
  {journal} {Phys. Rev. X}\ }\textbf {\bibinfo {volume} {7}},\ \bibinfo {pages}
  {011026} (\bibinfo {year} {2017})}\BibitemShut {NoStop}%
\bibitem [{\citenamefont {Medenjak}\ \emph {et~al.}(2020)\citenamefont
  {Medenjak}, \citenamefont {Bu\ifmmode~\check{c}\else \v{c}\fi{}a},\ and\
  \citenamefont {Jaksch}}]{Buca2020prb}%
  \BibitemOpen
  \bibfield  {author} {\bibinfo {author} {\bibfnamefont {M.}~\bibnamefont
  {Medenjak}}, \bibinfo {author} {\bibfnamefont {B.}~\bibnamefont
  {Bu\ifmmode~\check{c}\else \v{c}\fi{}a}},\ and\ \bibinfo {author}
  {\bibfnamefont {D.}~\bibnamefont {Jaksch}},\ }\bibfield  {title} {\bibinfo
  {title} {Isolated Heisenberg magnet as a quantum time crystal},\ }\href
  {https://doi.org/10.1103/PhysRevB.102.041117} {\bibfield  {journal} {\bibinfo
   {journal} {Phys. Rev. B}\ }\textbf {\bibinfo {volume} {102}},\ \bibinfo
  {pages} {041117(R)} (\bibinfo {year} {2020})}\BibitemShut {NoStop}%
\bibitem [{\citenamefont {Kongkhambut}\ \emph {et~al.}(2022)\citenamefont
  {Kongkhambut}, \citenamefont {Skulte}, \citenamefont {Mathey}, \citenamefont
  {Cosme}, \citenamefont {Hemmerich},\ and\ \citenamefont
  {Keßler}}]{Hemmerich2022Science}%
  \BibitemOpen
  \bibfield  {author} {\bibinfo {author} {\bibfnamefont {P.}~\bibnamefont
  {Kongkhambut}}, \bibinfo {author} {\bibfnamefont {J.}~\bibnamefont {Skulte}},
  \bibinfo {author} {\bibfnamefont {L.}~\bibnamefont {Mathey}}, \bibinfo
  {author} {\bibfnamefont {J.~G.}\ \bibnamefont {Cosme}}, \bibinfo {author}
  {\bibfnamefont {A.}~\bibnamefont {Hemmerich}},\ and\ \bibinfo {author}
  {\bibfnamefont {H.}~\bibnamefont {Keßler}},\ }\bibfield  {title} {\bibinfo
  {title} {Observation of a continuous time crystal},\ }\href
  {https://doi.org/10.1126/science.abo3382} {\bibfield  {journal} {\bibinfo
  {journal} {Science}\ }\textbf {\bibinfo {volume} {377}},\ \bibinfo {pages}
  {670} (\bibinfo {year} {2022})}\BibitemShut {NoStop}%
\bibitem [{\citenamefont {Liu}\ \emph {et~al.}(2023{\natexlab{a}})\citenamefont
  {Liu}, \citenamefont {Ou}, \citenamefont {MacDonald},\ and\ \citenamefont
  {Zheludev}}]{Liu2023NP}%
  \BibitemOpen
  \bibfield  {author} {\bibinfo {author} {\bibfnamefont {T.}~\bibnamefont
  {Liu}}, \bibinfo {author} {\bibfnamefont {J.-Y.}\ \bibnamefont {Ou}},
  \bibinfo {author} {\bibfnamefont {K.~F.}\ \bibnamefont {MacDonald}},\ and\
  \bibinfo {author} {\bibfnamefont {N.~I.}\ \bibnamefont {Zheludev}},\
  }\bibfield  {title} {\bibinfo {title} {Photonic metamaterial analogue of a
  continuous time crystal},\ }\href
  {https://doi.org/10.1038/s41567-023-02023-5} {\bibfield  {journal} {\bibinfo
  {journal} {Nat. Phys.}\ }\textbf {\bibinfo {volume} {19}},\ \bibinfo
  {pages} {986} (\bibinfo {year} {2023}{\natexlab{a}})}\BibitemShut
  {NoStop}%
\bibitem [{\citenamefont {Greilich}\ \emph {et~al.}(2024)\citenamefont
  {Greilich}, \citenamefont {Kopteva}, \citenamefont {Kamenskii}, \citenamefont
  {Sokolov}, \citenamefont {Korenev},\ and\ \citenamefont
  {Bayer}}]{Greilich2024NP}%
  \BibitemOpen
  \bibfield  {author} {\bibinfo {author} {\bibfnamefont {A.}~\bibnamefont
  {Greilich}}, \bibinfo {author} {\bibfnamefont {N.~E.}\ \bibnamefont
  {Kopteva}}, \bibinfo {author} {\bibfnamefont {A.~N.}\ \bibnamefont
  {Kamenskii}}, \bibinfo {author} {\bibfnamefont {P.~S.}\ \bibnamefont
  {Sokolov}}, \bibinfo {author} {\bibfnamefont {V.~L.}\ \bibnamefont
  {Korenev}},\ and\ \bibinfo {author} {\bibfnamefont {M.}~\bibnamefont
  {Bayer}},\ }\bibfield  {title} {\bibinfo {title} {Robust continuous time
  crystal in an electron–nuclear spin system},\ }\href
  {https://doi.org/10.1038/s41567-023-02351-6} {\bibfield  {journal} {\bibinfo
  {journal} {Nat. Phys.}\ }\textbf {\bibinfo {volume} {20}},\ \bibinfo
  {pages} {631} (\bibinfo {year} {2024})}\BibitemShut {NoStop}%
\bibitem [{\citenamefont {Wu}\ \emph {et~al.}(2024)\citenamefont {Wu},
  \citenamefont {Wang}, \citenamefont {Yang}, \citenamefont {Gao},
  \citenamefont {Liang}, \citenamefont {Tey}, \citenamefont {Li}, \citenamefont
  {Pohl},\ and\ \citenamefont {You}}]{You2024NP}%
  \BibitemOpen
  \bibfield  {author} {\bibinfo {author} {\bibfnamefont {X.}~\bibnamefont
  {Wu}}, \bibinfo {author} {\bibfnamefont {Z.}~\bibnamefont {Wang}}, \bibinfo
  {author} {\bibfnamefont {F.}~\bibnamefont {Yang}}, \bibinfo {author}
  {\bibfnamefont {R.}~\bibnamefont {Gao}}, \bibinfo {author} {\bibfnamefont
  {C.}~\bibnamefont {Liang}}, \bibinfo {author} {\bibfnamefont {M.~K.}\
  \bibnamefont {Tey}}, \bibinfo {author} {\bibfnamefont {X.}~\bibnamefont
  {Li}}, \bibinfo {author} {\bibfnamefont {T.}~\bibnamefont {Pohl}},\ and\
  \bibinfo {author} {\bibfnamefont {L.}~\bibnamefont {You}},\ }\bibfield
  {title} {\bibinfo {title} {Dissipative time crystal in a strongly interacting
  Rydberg gas},\ }\href {https://doi.org/10.1038/s41567-024-02542-9} {\bibfield
   {journal} {\bibinfo  {journal} {Nat. Phys.}\ }\textbf {\bibinfo {volume}
  {20}},\ \bibinfo {pages} {1389} (\bibinfo {year} {2024})}\BibitemShut
  {NoStop}%
\bibitem [{\citenamefont {Bull}\ \emph {et~al.}(2022)\citenamefont {Bull},
  \citenamefont {Hallam}, \citenamefont {Papi\ifmmode~\acute{c}\else
  \'{c}\fi{}},\ and\ \citenamefont {Martin}}]{CTCandQMBS2022}%
  \BibitemOpen
  \bibfield  {author} {\bibinfo {author} {\bibfnamefont {K.}~\bibnamefont
  {Bull}}, \bibinfo {author} {\bibfnamefont {A.}~\bibnamefont {Hallam}},
  \bibinfo {author} {\bibfnamefont {Z.}~\bibnamefont
  {Papi\ifmmode~\acute{c}\else \'{c}\fi{}}},\ and\ \bibinfo {author}
  {\bibfnamefont {I.}~\bibnamefont {Martin}},\ }\bibfield  {title} {\bibinfo
  {title} {Tuning between continuous time crystals and many-body scars in
  long-range $XYZ$ spin chains},\ }\href
  {https://doi.org/10.1103/PhysRevLett.129.140602} {\bibfield  {journal}
  {\bibinfo  {journal} {Phys. Rev. Lett.}\ }\textbf {\bibinfo {volume} {129}},\
  \bibinfo {pages} {140602} (\bibinfo {year} {2022})}\BibitemShut {NoStop}%
\bibitem [{\citenamefont {Bu\ifmmode~\check{c}\else
  \v{c}\fi{}a}(2023)}]{Buca2023prx}%
  \BibitemOpen
  \bibfield  {author} {\bibinfo {author} {\bibfnamefont {B.}~\bibnamefont
  {Bu\ifmmode~\check{c}\else \v{c}\fi{}a}},\ }\bibfield  {title} {\bibinfo
  {title} {Unified theory of local quantum many-body dynamics: Eigenoperator
  thermalization theorems},\ }\href
  {https://doi.org/10.1103/PhysRevX.13.031013} {\bibfield  {journal} {\bibinfo
  {journal} {Phys. Rev. X}\ }\textbf {\bibinfo {volume} {13}},\ \bibinfo
  {pages} {031013} (\bibinfo {year} {2023})}\BibitemShut {NoStop}%
\bibitem [{\citenamefont {Yang}\ and\ \citenamefont {Lee}(1952)}]{LeeYang1}%
  \BibitemOpen
  \bibfield  {author} {\bibinfo {author} {\bibfnamefont {C.~N.}\ \bibnamefont
  {Yang}}\ and\ \bibinfo {author} {\bibfnamefont {T.~D.}\ \bibnamefont {Lee}},\
  }\bibfield  {title} {\bibinfo {title} {Statistical theory of equations of
  state and phase transitions. I. Theory of condensation},\ }\href
  {https://doi.org/10.1103/PhysRev.87.404} {\bibfield  {journal} {\bibinfo
  {journal} {Phys. Rev.}\ }\textbf {\bibinfo {volume} {87}},\ \bibinfo {pages}
  {404} (\bibinfo {year} {1952})}\BibitemShut {NoStop}%
\bibitem [{\citenamefont {Lee}\ and\ \citenamefont {Yang}(1952)}]{LeeYang2}%
  \BibitemOpen
  \bibfield  {author} {\bibinfo {author} {\bibfnamefont {T.~D.}\ \bibnamefont
  {Lee}}\ and\ \bibinfo {author} {\bibfnamefont {C.~N.}\ \bibnamefont {Yang}},\
  }\bibfield  {title} {\bibinfo {title} {Statistical theory of equations of
  state and phase transitions. II. Lattice gas and Ising model},\ }\href
  {https://doi.org/10.1103/PhysRev.87.410} {\bibfield  {journal} {\bibinfo
  {journal} {Phys. Rev.}\ }\textbf {\bibinfo {volume} {87}},\ \bibinfo {pages}
  {410} (\bibinfo {year} {1952})}\BibitemShut {NoStop}%
\bibitem [{\citenamefont {Fisher}\ and\ \citenamefont
  {Brittin}(1965)}]{Fisher1965statistical}%
  \BibitemOpen
  \bibfield  {author} {\bibinfo {author} {\bibfnamefont {M.}~\bibnamefont
  {Fisher}}\ and\ \bibinfo {author} {\bibfnamefont {W.}~\bibnamefont
  {Brittin}},\ }\bibfield  {title} {\bibinfo {title} {\textit{Statistical Physics, Weak
  Interactions, Field Theory}},\ }\href@noop {} {\bibfield  {journal} {\bibinfo
  {journal} {Lectures in Theoretical Physics (University of Colorado
  Press, Boulder, 1965), Vol. VIIC}\ } }\BibitemShut {NoStop}%
\bibitem [{\citenamefont {Maldacena}\ \emph {et~al.}(2016)\citenamefont
  {Maldacena}, \citenamefont {Shenker},\ and\ \citenamefont
  {Stanford}}]{Maldacena2016jhep}%
  \BibitemOpen
  \bibfield  {author} {\bibinfo {author} {\bibfnamefont {J.}~\bibnamefont
  {Maldacena}}, \bibinfo {author} {\bibfnamefont {S.~H.}\ \bibnamefont
  {Shenker}},\ and\ \bibinfo {author} {\bibfnamefont {D.}~\bibnamefont
  {Stanford}},\ }\bibfield  {title} {\bibinfo {title} {A bound on chaos},\
  }\bibfield  {journal} {\bibinfo  {journal} {J. High Energy Phys.}\
  }\ \href{https://doi.org/10.1007/jhep08(2016)106}{08(2016)106}\BibitemShut {NoStop}%
\bibitem [{\citenamefont {Takahashi}\ and\ \citenamefont
  {Umezawa}(1996)}]{takahashi1996thermo}%
  \BibitemOpen
  \bibfield  {author} {\bibinfo {author} {\bibfnamefont {Y.}~\bibnamefont
  {Takahashi}}\ and\ \bibinfo {author} {\bibfnamefont {H.}~\bibnamefont
  {Umezawa}},\ }\bibfield  {title} {\bibinfo {title} {Thermo field dynamics},\
  }\href@noop {} {\bibfield  {journal} {\bibinfo  {journal} {Int. J. Mod. Phys. B}\ }\textbf {\bibinfo {volume} {10}},\ \bibinfo
  {pages} {1755} (\bibinfo {year} {1996})}\BibitemShut {NoStop}%
\bibitem [{\citenamefont {del Campo}\ \emph {et~al.}(2017)\citenamefont {del
  Campo}, \citenamefont {Molina-Vilaplana},\ and\ \citenamefont
  {Sonner}}]{delCampo2017PRD}%
  \BibitemOpen
  \bibfield  {author} {\bibinfo {author} {\bibfnamefont {A.}~\bibnamefont {del
  Campo}}, \bibinfo {author} {\bibfnamefont {J.}~\bibnamefont
  {Molina-Vilaplana}},\ and\ \bibinfo {author} {\bibfnamefont {J.}~\bibnamefont
  {Sonner}},\ }\bibfield  {title} {\bibinfo {title} {Scrambling the spectral
  form factor: Unitarity constraints and exact results},\ }\href
  {https://doi.org/10.1103/PhysRevD.95.126008} {\bibfield  {journal} {\bibinfo
  {journal} {Phys. Rev. D}\ }\textbf {\bibinfo {volume} {95}},\ \bibinfo
  {pages} {126008} (\bibinfo {year} {2017})}\BibitemShut {NoStop}%
\bibitem [{\citenamefont {Liu}\ \emph {et~al.}(2023{\natexlab{b}})\citenamefont
  {Liu}, \citenamefont {Lv}, \citenamefont {Yang},\ and\ \citenamefont
  {Zou}}]{Liu2023CPL}%
  \BibitemOpen
  \bibfield  {author} {\bibinfo {author} {\bibfnamefont {Y.}~\bibnamefont
  {Liu}}, \bibinfo {author} {\bibfnamefont {S.}~\bibnamefont {Lv}}, \bibinfo
  {author} {\bibfnamefont {Y.}~\bibnamefont {Yang}},\ and\ \bibinfo {author}
  {\bibfnamefont {H.}~\bibnamefont {Zou}},\ }\bibfield  {title} {\bibinfo
  {title} {Signatures of quantum criticality in the complex inverse temperature
  plane},\ }\href@noop {} {\bibfield  {journal} {\bibinfo  {journal} {Chin. Phys. Lett.}\ }\textbf {\bibinfo {volume} {40}},\ \bibinfo {pages}
  {050502} (\bibinfo {year} {2023}{\natexlab{b}})}\BibitemShut {NoStop}%
\bibitem [{\citenamefont {Liu}\ \emph {et~al.}(2024{\natexlab{a}})\citenamefont
  {Liu}, \citenamefont {Lv}, \citenamefont {Meng}, \citenamefont {Tan},
  \citenamefont {Zhao},\ and\ \citenamefont {Zou}}]{Liu2024PRR}%
  \BibitemOpen
  \bibfield  {author} {\bibinfo {author} {\bibfnamefont {Y.}~\bibnamefont
  {Liu}}, \bibinfo {author} {\bibfnamefont {S.}~\bibnamefont {Lv}}, \bibinfo
  {author} {\bibfnamefont {Y.}~\bibnamefont {Meng}}, \bibinfo {author}
  {\bibfnamefont {Z.}~\bibnamefont {Tan}}, \bibinfo {author} {\bibfnamefont
  {E.}~\bibnamefont {Zhao}},\ and\ \bibinfo {author} {\bibfnamefont
  {H.}~\bibnamefont {Zou}},\ }\bibfield  {title} {\bibinfo {title} {Exact
  fisher zeros and thermofield dynamics across a quantum critical point},\
  }\href {https://doi.org/10.1103/PhysRevResearch.6.043139} {\bibfield
  {journal} {\bibinfo  {journal} {Phys. Rev. Res.}\ }\textbf {\bibinfo {volume}
  {6}},\ \bibinfo {pages} {043139} (\bibinfo {year}
  {2024}{\natexlab{a}})}\BibitemShut {NoStop}%
\bibitem [{\citenamefont {Liu}\ \emph {et~al.}(2024{\natexlab{b}})\citenamefont
  {Liu}, \citenamefont {Zhao},\ and\ \citenamefont {Zou}}]{Liu2024CPL}%
  \BibitemOpen
  \bibfield  {author} {\bibinfo {author} {\bibfnamefont {Y.}~\bibnamefont
  {Liu}}, \bibinfo {author} {\bibfnamefont {E.}~\bibnamefont {Zhao}},\ and\
  \bibinfo {author} {\bibfnamefont {H.}~\bibnamefont {Zou}},\ }\bibfield
  {title} {\bibinfo {title} {From complexification to self-similarity: New
  aspects of quantum criticality},\ }\href
  {https://doi.org/10.1088/0256-307X/41/10/100501} {\bibfield  {journal}
  {\bibinfo  {journal} {Chin. Phys. Lett.}\ }\textbf {\bibinfo {volume}
  {41}},\ \bibinfo {pages} {100501} (\bibinfo {year}
  {2024}{\natexlab{b}})}\BibitemShut {NoStop}%
\bibitem [{\citenamefont {Heyl}\ \emph {et~al.}(2013)\citenamefont {Heyl},
  \citenamefont {Polkovnikov},\ and\ \citenamefont {Kehrein}}]{Heyl2013PRL}%
  \BibitemOpen
  \bibfield  {author} {\bibinfo {author} {\bibfnamefont {M.}~\bibnamefont
  {Heyl}}, \bibinfo {author} {\bibfnamefont {A.}~\bibnamefont {Polkovnikov}},\
  and\ \bibinfo {author} {\bibfnamefont {S.}~\bibnamefont {Kehrein}},\
  }\bibfield  {title} {\bibinfo {title} {Dynamical quantum phase transitions in
  the transverse-field Ising model},\ }\href
  {https://doi.org/10.1103/PhysRevLett.110.135704} {\bibfield  {journal}
  {\bibinfo  {journal} {Phys. Rev. Lett.}\ }\textbf {\bibinfo {volume} {110}},\
  \bibinfo {pages} {135704} (\bibinfo {year} {2013})}\BibitemShut {NoStop}%
\bibitem [{\citenamefont {Zvyagin}(2016)}]{reviewDPT2016}%
  \BibitemOpen
  \bibfield  {author} {\bibinfo {author} {\bibfnamefont {A.~A.}\ \bibnamefont
  {Zvyagin}},\ }\bibfield  {title} {\bibinfo {title} {Dynamical quantum phase
  transitions (review article)},\ }\href {https://doi.org/10.1063/1.4969869}
  {\bibfield  {journal} {\bibinfo  {journal} {Low Temp. Phys.}\
  }\textbf {\bibinfo {volume} {42}},\ \bibinfo {pages} {971} (\bibinfo {year}
  {2016})}\BibitemShut {NoStop}%
\bibitem [{\citenamefont {Heyl}(2018)}]{Heyl2018review}%
  \BibitemOpen
  \bibfield  {author} {\bibinfo {author} {\bibfnamefont {M.}~\bibnamefont
  {Heyl}},\ }\bibfield  {title} {\bibinfo {title} {Dynamical quantum phase
  transitions: A review},\ }\href {https://doi.org/10.1088/1361-6633/aaaf9a}
  {\bibfield  {journal} {\bibinfo  {journal} {Rep. Prog. Phys.}\
  }\textbf {\bibinfo {volume} {81}},\ \bibinfo {pages} {054001} (\bibinfo
  {year} {2018})}\BibitemShut {NoStop}%
\bibitem [{\citenamefont {Van~Damme}\ \emph {et~al.}(2023)\citenamefont
  {Van~Damme}, \citenamefont {Desaules}, \citenamefont
  {Papi\ifmmode~\acute{c}\else \'{c}\fi{}},\ and\ \citenamefont
  {Halimeh}}]{Halimeh2023prr}%
  \BibitemOpen
  \bibfield  {author} {\bibinfo {author} {\bibfnamefont {M.}~\bibnamefont
  {Van~Damme}}, \bibinfo {author} {\bibfnamefont {J.-Y.}\ \bibnamefont
  {Desaules}}, \bibinfo {author} {\bibfnamefont {Z.}~\bibnamefont
  {Papi\ifmmode~\acute{c}\else \'{c}\fi{}}},\ and\ \bibinfo {author}
  {\bibfnamefont {J.~C.}\ \bibnamefont {Halimeh}},\ }\bibfield  {title}
  {\bibinfo {title} {Anatomy of dynamical quantum phase transitions},\ }\href
  {https://doi.org/10.1103/PhysRevResearch.5.033090} {\bibfield  {journal}
  {\bibinfo  {journal} {Phys. Rev. Res.}\ }\textbf {\bibinfo {volume} {5}},\
  \bibinfo {pages} {033090} (\bibinfo {year} {2023})}\BibitemShut {NoStop}%
\bibitem [{\citenamefont {Lesanovsky}\ and\ \citenamefont
  {Katsura}(2012)}]{Lesanovsky2012pra}%
  \BibitemOpen
  \bibfield  {author} {\bibinfo {author} {\bibfnamefont {I.}~\bibnamefont
  {Lesanovsky}}\ and\ \bibinfo {author} {\bibfnamefont {H.}~\bibnamefont
  {Katsura}},\ }\bibfield  {title} {\bibinfo {title} {Interacting fibonacci
  anyons in a Rydberg gas},\ }\href
  {https://doi.org/10.1103/PhysRevA.86.041601} {\bibfield  {journal} {\bibinfo
  {journal} {Phys. Rev. A}\ }\textbf {\bibinfo {volume} {86}},\ \bibinfo
  {pages} {041601(R)} (\bibinfo {year} {2012})}\BibitemShut {NoStop}%
\bibitem [{\citenamefont {Turner}\ \emph
  {et~al.}(2018{\natexlab{b}})\citenamefont {Turner}, \citenamefont
  {Michailidis}, \citenamefont {Abanin}, \citenamefont {Serbyn},\ and\
  \citenamefont {Papi\'{c}}}]{Turner2018prb}%
  \BibitemOpen
  \bibfield  {author} {\bibinfo {author} {\bibfnamefont {C.~J.}\ \bibnamefont
  {Turner}}, \bibinfo {author} {\bibfnamefont {A.~A.}\ \bibnamefont
  {Michailidis}}, \bibinfo {author} {\bibfnamefont {D.~A.}\ \bibnamefont
  {Abanin}}, \bibinfo {author} {\bibfnamefont {M.}~\bibnamefont {Serbyn}},\
  and\ \bibinfo {author} {\bibfnamefont {Z.}~\bibnamefont {Papi\'{c}}},\
  }\bibfield  {title} {\bibinfo {title} {Quantum scarred eigenstates in a
  Rydberg atom chain: Entanglement, breakdown of thermalization, and stability
  to perturbations},\ }\href {https://doi.org/10.1103/PhysRevB.98.155134}
  {\bibfield  {journal} {\bibinfo  {journal} {Phys. Rev. B}\ }\textbf {\bibinfo
  {volume} {98}},\ \bibinfo {pages} {155134} (\bibinfo {year}
  {2018}{\natexlab{b}})}\BibitemShut {NoStop}%
\bibitem [{\citenamefont {Ho}\ \emph {et~al.}(2019)\citenamefont {Ho},
  \citenamefont {Choi}, \citenamefont {Pichler},\ and\ \citenamefont
  {Lukin}}]{Ho2019prl}%
  \BibitemOpen
  \bibfield  {author} {\bibinfo {author} {\bibfnamefont {W.~W.}\ \bibnamefont
  {Ho}}, \bibinfo {author} {\bibfnamefont {S.}~\bibnamefont {Choi}}, \bibinfo
  {author} {\bibfnamefont {H.}~\bibnamefont {Pichler}},\ and\ \bibinfo {author}
  {\bibfnamefont {M.~D.}\ \bibnamefont {Lukin}},\ }\bibfield  {title} {\bibinfo
  {title} {Periodic orbits, entanglement, and quantum many-body scars in
  constrained models: Matrix product state approach},\ }\href
  {https://doi.org/10.1103/PhysRevLett.122.040603} {\bibfield  {journal}
  {\bibinfo  {journal} {Phys. Rev. Lett.}\ }\textbf {\bibinfo {volume} {122}},\
  \bibinfo {pages} {040603} (\bibinfo {year} {2019})}\BibitemShut {NoStop}%
\bibitem [{\citenamefont {Zhang}\ \emph {et~al.}(2023)\citenamefont {Zhang},
  \citenamefont {Yuan}, \citenamefont {Iadecola}, \citenamefont {Xu},\ and\
  \citenamefont {Deng}}]{Deng2023prl}%
  \BibitemOpen
  \bibfield  {author} {\bibinfo {author} {\bibfnamefont {S.-Y.}\ \bibnamefont
  {Zhang}}, \bibinfo {author} {\bibfnamefont {D.}~\bibnamefont {Yuan}},
  \bibinfo {author} {\bibfnamefont {T.}~\bibnamefont {Iadecola}}, \bibinfo
  {author} {\bibfnamefont {S.}~\bibnamefont {Xu}},\ and\ \bibinfo {author}
  {\bibfnamefont {D.-L.}\ \bibnamefont {Deng}},\ }\bibfield  {title} {\bibinfo
  {title} {Extracting quantum many-body scarred eigenstates with matrix product
  states},\ }\href {https://doi.org/10.1103/PhysRevLett.131.020402} {\bibfield
  {journal} {\bibinfo  {journal} {Phys. Rev. Lett.}\ }\textbf {\bibinfo
  {volume} {131}},\ \bibinfo {pages} {020402} (\bibinfo {year}
  {2023})}\BibitemShut {NoStop}%
\bibitem [{\citenamefont {Ivanov}\ and\ \citenamefont
  {Motrunich}(2025)}]{Ivanov2025arxiv}%
  \BibitemOpen
  \bibfield  {author} {\bibinfo {author} {\bibfnamefont {A.}~\bibnamefont
  {Ivanov}}\ and\ \bibinfo {author} {\bibfnamefont {O.}~\bibnamefont
  {Motrunich}},\ }\href@noop {} {\bibinfo {title} {Many exact area-law scar
  eigenstates in the nonintegrable PXP and related models}},\ \Eprint {https://arxiv.org/abs/arXiv:2503.16327} {arXiv:2503.16327}
  \BibitemShut {NoStop}%
\bibitem [{\citenamefont {Or\'us}(2014)}]{ORUS2014117}%
  \BibitemOpen
  \bibfield  {author} {\bibinfo {author} {\bibfnamefont {R.}~\bibnamefont
  {Or\'us}},\ }\bibfield  {title} {\bibinfo {title} {A practical introduction
  to tensor networks: Matrix product states and projected entangled pair
  states},\ }\href {https://doi.org/https://doi.org/10.1016/j.aop.2014.06.013}
  {\bibfield  {journal} {\bibinfo  {journal} {Ann. Phys. (Amsterdam)}\ }\textbf
  {\bibinfo {volume} {349}},\ \bibinfo {pages} {117} (\bibinfo {year}
  {2014})}\BibitemShut {NoStop}%
\bibitem [{\citenamefont {Cirac}\ \emph {et~al.}(2021)\citenamefont {Cirac},
  \citenamefont {P\'erez-Garc\'{\i}a}, \citenamefont {Schuch},\ and\
  \citenamefont {Verstraete}}]{TNreview1}%
  \BibitemOpen
  \bibfield  {author} {\bibinfo {author} {\bibfnamefont {J.~I.}\ \bibnamefont
  {Cirac}}, \bibinfo {author} {\bibfnamefont {D.}~\bibnamefont
  {P\'erez-Garc\'{\i}a}}, \bibinfo {author} {\bibfnamefont {N.}~\bibnamefont
  {Schuch}},\ and\ \bibinfo {author} {\bibfnamefont {F.}~\bibnamefont
  {Verstraete}},\ }\bibfield  {title} {\bibinfo {title} {Matrix product states
  and projected entangled pair states: Concepts, symmetries, theorems},\ }\href
  {https://doi.org/10.1103/RevModPhys.93.045003} {\bibfield  {journal}
  {\bibinfo  {journal} {Rev. Mod. Phys.}\ }\textbf {\bibinfo {volume} {93}},\
  \bibinfo {pages} {045003} (\bibinfo {year} {2021})}\BibitemShut {NoStop}%
\bibitem [{\citenamefont {Meurice}\ \emph {et~al.}(2022)\citenamefont
  {Meurice}, \citenamefont {Sakai},\ and\ \citenamefont
  {Unmuth-Yockey}}]{TNreview2}%
  \BibitemOpen
  \bibfield  {author} {\bibinfo {author} {\bibfnamefont {Y.}~\bibnamefont
  {Meurice}}, \bibinfo {author} {\bibfnamefont {R.}~\bibnamefont {Sakai}},\
  and\ \bibinfo {author} {\bibfnamefont {J.}~\bibnamefont {Unmuth-Yockey}},\
  }\bibfield  {title} {\bibinfo {title} {Tensor lattice field theory for
  renormalization and quantum computing},\ }\href
  {https://doi.org/10.1103/RevModPhys.94.025005} {\bibfield  {journal}
  {\bibinfo  {journal} {Rev. Mod. Phys.}\ }\textbf {\bibinfo {volume} {94}},\
  \bibinfo {pages} {025005} (\bibinfo {year} {2022})}\BibitemShut {NoStop}%
\bibitem [{\citenamefont {Suzuki}(1976)}]{Suzuki1976}%
  \BibitemOpen
  \bibfield  {author} {\bibinfo {author} {\bibfnamefont {M.}~\bibnamefont
  {Suzuki}},\ }\bibfield  {title} {\bibinfo {title} {Relationship between
  d-dimensional quantal spin systems and (d$+$1)-dimensional Ising systems:
  Equivalence, critical exponents and systematic approximants of the partition
  function and spin correlations},\ }\href
  {https://doi.org/10.1143/ptp.56.1454} {\bibfield  {journal} {\bibinfo
  {journal} {Prog. Theor. Phys.}\ }\textbf {\bibinfo {volume}
  {56}},\ \bibinfo {pages} {1454} (\bibinfo {year} {1976})}\BibitemShut
  {NoStop}%
\bibitem [{\citenamefont {Xie}\ \emph {et~al.}(2012)\citenamefont {Xie},
  \citenamefont {Chen}, \citenamefont {Qin}, \citenamefont {Zhu}, \citenamefont
  {Yang},\ and\ \citenamefont {Xiang}}]{XieHOTRG}%
  \BibitemOpen
  \bibfield  {author} {\bibinfo {author} {\bibfnamefont {Z.~Y.}\ \bibnamefont
  {Xie}}, \bibinfo {author} {\bibfnamefont {J.}~\bibnamefont {Chen}}, \bibinfo
  {author} {\bibfnamefont {M.~P.}\ \bibnamefont {Qin}}, \bibinfo {author}
  {\bibfnamefont {J.~W.}\ \bibnamefont {Zhu}}, \bibinfo {author} {\bibfnamefont
  {L.~P.}\ \bibnamefont {Yang}},\ and\ \bibinfo {author} {\bibfnamefont
  {T.}~\bibnamefont {Xiang}},\ }\bibfield  {title} {\bibinfo {title}
  {Coarse-graining renormalization by higher-order singular value
  decomposition},\ }\href {https://doi.org/10.1103/PhysRevB.86.045139}
  {\bibfield  {journal} {\bibinfo  {journal} {Phys. Rev. B}\ }\textbf {\bibinfo
  {volume} {86}},\ \bibinfo {pages} {045139} (\bibinfo {year}
  {2012})}\BibitemShut {NoStop}%
\bibitem [{\citenamefont {Denbleyker}\ \emph {et~al.}(2014)\citenamefont
  {Denbleyker}, \citenamefont {Liu}, \citenamefont {Meurice}, \citenamefont
  {Qin}, \citenamefont {Xiang}, \citenamefont {Xie}, \citenamefont {Yu},\ and\
  \citenamefont {Zou}}]{Zou2014PRD}%
  \BibitemOpen
  \bibfield  {author} {\bibinfo {author} {\bibfnamefont {A.}~\bibnamefont
  {Denbleyker}}, \bibinfo {author} {\bibfnamefont {Y.}~\bibnamefont {Liu}},
  \bibinfo {author} {\bibfnamefont {Y.}~\bibnamefont {Meurice}}, \bibinfo
  {author} {\bibfnamefont {M.~P.}\ \bibnamefont {Qin}}, \bibinfo {author}
  {\bibfnamefont {T.}~\bibnamefont {Xiang}}, \bibinfo {author} {\bibfnamefont
  {Z.~Y.}\ \bibnamefont {Xie}}, \bibinfo {author} {\bibfnamefont {J.~F.}\
  \bibnamefont {Yu}},\ and\ \bibinfo {author} {\bibfnamefont {H.}~\bibnamefont
  {Zou}},\ }\bibfield  {title} {\bibinfo {title} {Controlling sign problems in
  spin models using tensor renormalization},\ }\href
  {https://doi.org/10.1103/PhysRevD.89.016008} {\bibfield  {journal} {\bibinfo
  {journal} {Phys. Rev. D}\ }\textbf {\bibinfo {volume} {89}},\ \bibinfo
  {pages} {016008} (\bibinfo {year} {2014})}\BibitemShut {NoStop}%
\bibitem [{sup()}]{supp}%
  \BibitemOpen
  \href@noop {} {}\bibinfo {note} {See Supplemental Material at http://link.aps.org/supplemental/10.1103/glc5-hv2m for the
  details on (a) Fisher zeros of the $\bar{P}X\bar{P}$ model and the
  fixed-point line, (b) Fisher zeros of $Z_i$ in the $\bar{P}X\bar{P}$ model,
  (c) Fisher zeros of a lattice gauge field related model, (d) the Ising model
  with external fields. The Supplemental Material also contains
  Refs.[36,37,55-59]}\BibitemShut {NoStop}%
\bibitem [{\citenamefont {Borla}\ \emph {et~al.}(2020)\citenamefont {Borla},
  \citenamefont {Verresen}, \citenamefont {Grusdt},\ and\ \citenamefont
  {Moroz}}]{Sergej2020prl}%
  \BibitemOpen
  \bibfield  {author} {\bibinfo {author} {\bibfnamefont {U.}~\bibnamefont
  {Borla}}, \bibinfo {author} {\bibfnamefont {R.}~\bibnamefont {Verresen}},
  \bibinfo {author} {\bibfnamefont {F.}~\bibnamefont {Grusdt}},\ and\ \bibinfo
  {author} {\bibfnamefont {S.}~\bibnamefont {Moroz}},\ }\bibfield  {title}
  {\bibinfo {title} {Confined phases of one-dimensional spinless fermions
  coupled to ${Z}_{2}$ gauge theory},\ }\href
  {https://doi.org/10.1103/PhysRevLett.124.120503} {\bibfield  {journal}
  {\bibinfo  {journal} {Phys. Rev. Lett.}\ }\textbf {\bibinfo {volume} {124}},\
  \bibinfo {pages} {120503} (\bibinfo {year} {2020})}\BibitemShut {NoStop}%
\bibitem [{\citenamefont {Iadecola}\ and\ \citenamefont
  {Schecter}(2020)}]{Iadecola2020prb}%
  \BibitemOpen
  \bibfield  {author} {\bibinfo {author} {\bibfnamefont {T.}~\bibnamefont
  {Iadecola}}\ and\ \bibinfo {author} {\bibfnamefont {M.}~\bibnamefont
  {Schecter}},\ }\bibfield  {title} {\bibinfo {title} {Quantum many-body scar
  states with emergent kinetic constraints and finite-entanglement revivals},\
  }\href {https://doi.org/10.1103/PhysRevB.101.024306} {\bibfield  {journal}
  {\bibinfo  {journal} {Phys. Rev. B}\ }\textbf {\bibinfo {volume} {101}},\
  \bibinfo {pages} {024306} (\bibinfo {year} {2020})}\BibitemShut {NoStop}%
\bibitem [{\citenamefont {Halimeh}\ \emph {et~al.}(2023)\citenamefont
  {Halimeh}, \citenamefont {Barbiero}, \citenamefont {Hauke}, \citenamefont
  {Grusdt},\ and\ \citenamefont {Bohrdt}}]{Halimeh2023Quantum}%
  \BibitemOpen
  \bibfield  {author} {\bibinfo {author} {\bibfnamefont {J.~C.}\ \bibnamefont
  {Halimeh}}, \bibinfo {author} {\bibfnamefont {L.}~\bibnamefont {Barbiero}},
  \bibinfo {author} {\bibfnamefont {P.}~\bibnamefont {Hauke}}, \bibinfo
  {author} {\bibfnamefont {F.}~\bibnamefont {Grusdt}},\ and\ \bibinfo {author}
  {\bibfnamefont {A.}~\bibnamefont {Bohrdt}},\ }\bibfield  {title} {\bibinfo
  {title} {Robust quantum many-body scars in lattice gauge theories},\ }\href
  {https://doi.org/10.22331/q-2023-05-15-1004} {\bibfield  {journal} {\bibinfo
  {journal} {{Quantum}}\ }\textbf {\bibinfo {volume} {7}},\ \bibinfo {pages}
  {1004} (\bibinfo {year} {2023})}\BibitemShut {NoStop}%
\bibitem [{\citenamefont {Denbleyker}\ \emph {et~al.}(2010)\citenamefont
  {Denbleyker}, \citenamefont {Du}, \citenamefont {Liu}, \citenamefont
  {Meurice},\ and\ \citenamefont {Zou}}]{RGflow2010}%
  \BibitemOpen
  \bibfield  {author} {\bibinfo {author} {\bibfnamefont {A.}~\bibnamefont
  {Denbleyker}}, \bibinfo {author} {\bibfnamefont {D.}~\bibnamefont {Du}},
  \bibinfo {author} {\bibfnamefont {Y.}~\bibnamefont {Liu}}, \bibinfo {author}
  {\bibfnamefont {Y.}~\bibnamefont {Meurice}},\ and\ \bibinfo {author}
  {\bibfnamefont {H.}~\bibnamefont {Zou}},\ }\bibfield  {title} {\bibinfo
  {title} {Fisher's zeros as the boundary of renormalization group flows in
  complex coupling spaces},\ }\href
  {https://doi.org/10.1103/PhysRevLett.104.251601} {\bibfield  {journal}
  {\bibinfo  {journal} {Phys. Rev. Lett.}\ }\textbf {\bibinfo {volume} {104}},\
  \bibinfo {pages} {251601} (\bibinfo {year} {2010})}\BibitemShut {NoStop}%
\bibitem [{\citenamefont {Meurice}\ and\ \citenamefont
  {Zou}(2011)}]{Zou2011PRD}%
  \BibitemOpen
  \bibfield  {author} {\bibinfo {author} {\bibfnamefont {Y.}~\bibnamefont
  {Meurice}}\ and\ \bibinfo {author} {\bibfnamefont {H.}~\bibnamefont {Zou}},\
  }\bibfield  {title} {\bibinfo {title} {Complex renormalization group flows
  for 2d nonlinear $o(n)$ sigma models},\ }\href
  {https://doi.org/10.1103/PhysRevD.83.056009} {\bibfield  {journal} {\bibinfo
  {journal} {Phys. Rev. D}\ }\textbf {\bibinfo {volume} {83}},\ \bibinfo
  {pages} {056009} (\bibinfo {year} {2011})}\BibitemShut {NoStop}%
\bibitem [{\citenamefont {Affleck}\ \emph {et~al.}(1987)\citenamefont
  {Affleck}, \citenamefont {Kennedy}, \citenamefont {Lieb},\ and\ \citenamefont
  {Tasaki}}]{AKLT1987}%
  \BibitemOpen
  \bibfield  {author} {\bibinfo {author} {\bibfnamefont {I.}~\bibnamefont
  {Affleck}}, \bibinfo {author} {\bibfnamefont {T.}~\bibnamefont {Kennedy}},
  \bibinfo {author} {\bibfnamefont {E.~H.}\ \bibnamefont {Lieb}},\ and\
  \bibinfo {author} {\bibfnamefont {H.}~\bibnamefont {Tasaki}},\ }\bibfield
  {title} {\bibinfo {title} {Rigorous results on valence-bond ground states in
  antiferromagnets},\ }\href {https://doi.org/10.1103/PhysRevLett.59.799}
  {\bibfield  {journal} {\bibinfo  {journal} {Phys. Rev. Lett.}\ }\textbf
  {\bibinfo {volume} {59}},\ \bibinfo {pages} {799} (\bibinfo {year}
  {1987})}\BibitemShut {NoStop}%
\bibitem [{\citenamefont {Shen}\ \emph {et~al.}(2023)\citenamefont {Shen},
  \citenamefont {Guo},\ and\ \citenamefont {Yang}}]{Yang2023Construction}%
  \BibitemOpen
  \bibfield  {author} {\bibinfo {author} {\bibfnamefont {R.}~\bibnamefont
  {Shen}}, \bibinfo {author} {\bibfnamefont {Y.}~\bibnamefont {Guo}},\ and\
  \bibinfo {author} {\bibfnamefont {S.}~\bibnamefont {Yang}},\ }\bibfield
  {title} {\bibinfo {title} {Construction of non-Hermitian parent Hamiltonian
  from matrix product states},\ }\href
  {https://doi.org/10.1103/PhysRevLett.130.220401} {\bibfield  {journal}
  {\bibinfo  {journal} {Phys. Rev. Lett.}\ }\textbf {\bibinfo {volume} {130}},\
  \bibinfo {pages} {220401} (\bibinfo {year} {2023})}\BibitemShut {NoStop}%
\bibitem [{Zol()}]{Zollar2024}%
  \BibitemOpen
  \href@noop {} {}\bibinfo {note} {Peter Zoller, Closing remarks at the 2nd
  Conference on `Quantum Simulations of Fundamental Physics'
  (https://m.koushare.com/live/details/37543) and the International Workshop on
  `Quantum Systems with Novel Spatiotemporal Control'
  (https://m.koushare.com/live/details/38269).}\BibitemShut {Stop}%
\bibitem [{\citenamefont {Francis}\ \emph {et~al.}(2021)\citenamefont
  {Francis}, \citenamefont {Zhu}, \citenamefont {Huerta~Alderete},
  \citenamefont {Johri}, \citenamefont {Xiao}, \citenamefont {Freericks},
  \citenamefont {Monroe}, \citenamefont {Linke},\ and\ \citenamefont
  {Kemper}}]{Francis2021SA}%
  \BibitemOpen
  \bibfield  {author} {\bibinfo {author} {\bibfnamefont {A.}~\bibnamefont
  {Francis}}, \bibinfo {author} {\bibfnamefont {D.}~\bibnamefont {Zhu}},
  \bibinfo {author} {\bibfnamefont {C.}~\bibnamefont {Huerta~Alderete}},
  \bibinfo {author} {\bibfnamefont {S.}~\bibnamefont {Johri}}, \bibinfo
  {author} {\bibfnamefont {X.}~\bibnamefont {Xiao}}, \bibinfo {author}
  {\bibfnamefont {J.~K.}\ \bibnamefont {Freericks}}, \bibinfo {author}
  {\bibfnamefont {C.}~\bibnamefont {Monroe}}, \bibinfo {author} {\bibfnamefont
  {N.~M.}\ \bibnamefont {Linke}},\ and\ \bibinfo {author} {\bibfnamefont
  {A.~F.}\ \bibnamefont {Kemper}},\ }\bibfield  {title} {\bibinfo {title}
  {Many-body thermodynamics on quantum computers via partition function
  zeros},\ }\bibfield  {journal} {\bibinfo  {journal} {Sci. Adv.}\
  }\textbf {\bibinfo {volume} {7}},\ \href
  {https://doi.org/10.1126/sciadv.abf2447} {10.1126/sciadv.abf2447} (\bibinfo
  {year} {2021})\BibitemShut {NoStop}%
\bibitem [{\citenamefont {Fisher}\ \emph {et~al.}(2023)\citenamefont {Fisher},
  \citenamefont {Khemani}, \citenamefont {Nahum},\ and\ \citenamefont
  {Vijay}}]{fisher2023random}%
  \BibitemOpen
  \bibfield  {author} {\bibinfo {author} {\bibfnamefont {M.~P.}\ \bibnamefont
  {Fisher}}, \bibinfo {author} {\bibfnamefont {V.}~\bibnamefont {Khemani}},
  \bibinfo {author} {\bibfnamefont {A.}~\bibnamefont {Nahum}},\ and\ \bibinfo
  {author} {\bibfnamefont {S.}~\bibnamefont {Vijay}},\ }\bibfield  {title}
  {\bibinfo {title} {Random quantum circuits},\ }\href@noop {} {\bibfield
  {journal} {\bibinfo  {journal} {Annu. Rev. Condens. Matter Phys.}\
  }\textbf {\bibinfo {volume} {14}},\ \bibinfo {pages} {335} (\bibinfo {year}
  {2023})}\BibitemShut {NoStop}%
\bibitem [{\citenamefont {Granet}\ \emph {et~al.}(2023)\citenamefont {Granet},
  \citenamefont {Zhang},\ and\ \citenamefont {Dreyer}}]{Granet2023prl}%
  \BibitemOpen
  \bibfield  {author} {\bibinfo {author} {\bibfnamefont {E.}~\bibnamefont
  {Granet}}, \bibinfo {author} {\bibfnamefont {C.}~\bibnamefont {Zhang}},\ and\
  \bibinfo {author} {\bibfnamefont {H.}~\bibnamefont {Dreyer}},\ }\bibfield
  {title} {\bibinfo {title} {Volume-law to area-law entanglement transition in
  a nonunitary periodic Gaussian circuit},\ }\href
  {https://doi.org/10.1103/PhysRevLett.130.230401} {\bibfield  {journal}
  {\bibinfo  {journal} {Phys. Rev. Lett.}\ }\textbf {\bibinfo {volume} {130}},\
  \bibinfo {pages} {230401} (\bibinfo {year} {2023})}\BibitemShut {NoStop}%
\bibitem [{\citenamefont {Peng}\ \emph {et~al.}(2015)\citenamefont {Peng},
  \citenamefont {Zhou}, \citenamefont {Wei}, \citenamefont {Cui}, \citenamefont
  {Du},\ and\ \citenamefont {Liu}}]{experimentLeeYang}%
  \BibitemOpen
  \bibfield  {author} {\bibinfo {author} {\bibfnamefont {X.}~\bibnamefont
  {Peng}}, \bibinfo {author} {\bibfnamefont {H.}~\bibnamefont {Zhou}}, \bibinfo
  {author} {\bibfnamefont {B.-B.}\ \bibnamefont {Wei}}, \bibinfo {author}
  {\bibfnamefont {J.}~\bibnamefont {Cui}}, \bibinfo {author} {\bibfnamefont
  {J.}~\bibnamefont {Du}},\ and\ \bibinfo {author} {\bibfnamefont {R.-B.}\
  \bibnamefont {Liu}},\ }\bibfield  {title} {\bibinfo {title} {Experimental
  observation of Lee-Yang zeros},\ }\href
  {https://doi.org/10.1103/PhysRevLett.114.010601} {\bibfield  {journal}
  {\bibinfo  {journal} {Phys. Rev. Lett.}\ }\textbf {\bibinfo {volume} {114}},\
  \bibinfo {pages} {010601} (\bibinfo {year} {2015})}\BibitemShut {NoStop}%
\bibitem [{\citenamefont {Meng}\ \emph {et~al.}(2025)\citenamefont {Meng},
  \citenamefont {Lv}, \citenamefont {Liu}, \citenamefont {Tan}, \citenamefont
  {Zhao},\ and\ \citenamefont {Zou}}]{figdata}%
  \BibitemOpen
  \bibfield  {author} {\bibinfo {author} {\bibfnamefont {Y.}~\bibnamefont
  {Meng}}, \bibinfo {author} {\bibfnamefont {S.}~\bibnamefont {Lv}}, \bibinfo
  {author} {\bibfnamefont {Y.}~\bibnamefont {Liu}}, \bibinfo {author}
  {\bibfnamefont {Z.}~\bibnamefont {Tan}}, \bibinfo {author} {\bibfnamefont
  {E.}~\bibnamefont {Zhao}},\ and\ \bibinfo {author} {\bibfnamefont
  {H.}~\bibnamefont {Zou}},\ }\href {https://doi.org/10.5281/ZENODO.16616679}
  {\bibinfo {title} {Figure data for detecting many-body scars from Fisher
  zeros (https://doi.org/10.5281/zenodo.16616679)}} (\bibinfo {year} {2025})\BibitemShut {NoStop}%

\end{thebibliography}

\begin{thebibliography}{7}%
\makeatletter
\providecommand \@ifxundefined [1]{%
 \@ifx{#1\undefined}
}%
\providecommand \@ifnum [1]{%
 \ifnum #1\expandafter \@firstoftwo
 \else \expandafter \@secondoftwo
 \fi
}%
\providecommand \@ifx [1]{%
 \ifx #1\expandafter \@firstoftwo
 \else \expandafter \@secondoftwo
 \fi
}%
\providecommand \natexlab [1]{#1}%
\providecommand \enquote  [1]{``#1''}%
\providecommand \bibnamefont  [1]{#1}%
\providecommand \bibfnamefont [1]{#1}%
\providecommand \citenamefont [1]{#1}%
\providecommand \href@noop [0]{\@secondoftwo}%
\providecommand \href [0]{\begingroup \@sanitize@url \@href}%
\providecommand \@href[1]{\@@startlink{#1}\@@href}%
\providecommand \@@href[1]{\endgroup#1\@@endlink}%
\providecommand \@sanitize@url [0]{\catcode `\\12\catcode `\$12\catcode
  `\&12\catcode `\#12\catcode `\^12\catcode `\_12\catcode `\%12\relax}%
\providecommand \@@startlink[1]{}%
\providecommand \@@endlink[0]{}%
\providecommand \url  [0]{\begingroup\@sanitize@url \@url }%
\providecommand \@url [1]{\endgroup\@href {#1}{\urlprefix }}%
\providecommand \urlprefix  [0]{URL }%
\providecommand \Eprint [0]{\href }%
\providecommand \doibase [0]{https://doi.org/}%
\providecommand \selectlanguage [0]{\@gobble}%
\providecommand \bibinfo  [0]{\@secondoftwo}%
\providecommand \bibfield  [0]{\@secondoftwo}%
\providecommand \translation [1]{[#1]}%
\providecommand \BibitemOpen [0]{}%
\providecommand \bibitemStop [0]{}%
\providecommand \bibitemNoStop [0]{.\EOS\space}%
\providecommand \EOS [0]{\spacefactor3000\relax}%
\providecommand \BibitemShut  [1]{\csname bibitem#1\endcsname}%
\let\auto@bib@innerbib\@empty
\bibitem [{\citenamefont {Liu}\ \emph {et~al.}(2024)\citenamefont {Liu},
  \citenamefont {Lv}, \citenamefont {Meng}, \citenamefont {Tan}, \citenamefont
  {Zhao},\ and\ \citenamefont {Zou}}]{SLiu2024PRR}%
  \BibitemOpen
  \bibfield  {author} {\bibinfo {author} {\bibfnamefont {Y.}~\bibnamefont
  {Liu}}, \bibinfo {author} {\bibfnamefont {S.}~\bibnamefont {Lv}}, \bibinfo
  {author} {\bibfnamefont {Y.}~\bibnamefont {Meng}}, \bibinfo {author}
  {\bibfnamefont {Z.}~\bibnamefont {Tan}}, \bibinfo {author} {\bibfnamefont
  {E.}~\bibnamefont {Zhao}},\ and\ \bibinfo {author} {\bibfnamefont
  {H.}~\bibnamefont {Zou}},\ }\bibfield  {title} {\bibinfo {title} {Exact
  fisher zeros and thermofield dynamics across a quantum critical point},\
  }\href {https://doi.org/10.1103/PhysRevResearch.6.043139} {\bibfield
  {journal} {\bibinfo  {journal} {Phys. Rev. Res.}\ }\textbf {\bibinfo {volume}
  {6}},\ \bibinfo {pages} {043139} (\bibinfo {year} {2024})}\BibitemShut
  {NoStop}%
\bibitem [{\citenamefont {Denbleyker}\ \emph {et~al.}(2010)\citenamefont
  {Denbleyker}, \citenamefont {Du}, \citenamefont {Liu}, \citenamefont
  {Meurice},\ and\ \citenamefont {Zou}}]{SRGflow2010}%
  \BibitemOpen
  \bibfield  {author} {\bibinfo {author} {\bibfnamefont {A.}~\bibnamefont
  {Denbleyker}}, \bibinfo {author} {\bibfnamefont {D.}~\bibnamefont {Du}},
  \bibinfo {author} {\bibfnamefont {Y.}~\bibnamefont {Liu}}, \bibinfo {author}
  {\bibfnamefont {Y.}~\bibnamefont {Meurice}},\ and\ \bibinfo {author}
  {\bibfnamefont {H.}~\bibnamefont {Zou}},\ }\bibfield  {title} {\bibinfo
  {title} {Fisher's zeros as the boundary of renormalization group flows in
  complex coupling spaces},\ }\href
  {https://doi.org/10.1103/PhysRevLett.104.251601} {\bibfield  {journal}
  {\bibinfo  {journal} {Phys. Rev. Lett.}\ }\textbf {\bibinfo {volume} {104}},\
  \bibinfo {pages} {251601} (\bibinfo {year} {2010})}\BibitemShut {NoStop}%
\bibitem [{\citenamefont {Meurice}\ and\ \citenamefont
  {Zou}(2011)}]{SZou2011PRD}%
  \BibitemOpen
  \bibfield  {author} {\bibinfo {author} {\bibfnamefont {Y.}~\bibnamefont
  {Meurice}}\ and\ \bibinfo {author} {\bibfnamefont {H.}~\bibnamefont {Zou}},\
  }\bibfield  {title} {\bibinfo {title} {Complex renormalization group flows
  for 2d nonlinear $o(n)$ sigma models},\ }\href
  {https://doi.org/10.1103/PhysRevD.83.056009} {\bibfield  {journal} {\bibinfo
  {journal} {Phys. Rev. D}\ }\textbf {\bibinfo {volume} {83}},\ \bibinfo
  {pages} {056009} (\bibinfo {year} {2011})}\BibitemShut {NoStop}%
\bibitem [{\citenamefont {Borla}\ \emph {et~al.}(2020)\citenamefont {Borla},
  \citenamefont {Verresen}, \citenamefont {Grusdt},\ and\ \citenamefont
  {Moroz}}]{SSergej2020prl}%
  \BibitemOpen
  \bibfield  {author} {\bibinfo {author} {\bibfnamefont {U.}~\bibnamefont
  {Borla}}, \bibinfo {author} {\bibfnamefont {R.}~\bibnamefont {Verresen}},
  \bibinfo {author} {\bibfnamefont {F.}~\bibnamefont {Grusdt}},\ and\ \bibinfo
  {author} {\bibfnamefont {S.}~\bibnamefont {Moroz}},\ }\bibfield  {title}
  {\bibinfo {title} {Confined phases of one-dimensional spinless fermions
  coupled to ${Z}_{2}$ gauge theory},\ }\href
  {https://doi.org/10.1103/PhysRevLett.124.120503} {\bibfield  {journal}
  {\bibinfo  {journal} {Phys. Rev. Lett.}\ }\textbf {\bibinfo {volume} {124}},\
  \bibinfo {pages} {120503} (\bibinfo {year} {2020})}\BibitemShut {NoStop}%
\bibitem [{\citenamefont {Iadecola}\ and\ \citenamefont
  {Schecter}(2020)}]{SIadecola2020prb}%
  \BibitemOpen
  \bibfield  {author} {\bibinfo {author} {\bibfnamefont {T.}~\bibnamefont
  {Iadecola}}\ and\ \bibinfo {author} {\bibfnamefont {M.}~\bibnamefont
  {Schecter}},\ }\bibfield  {title} {\bibinfo {title} {Quantum many-body scar
  states with emergent kinetic constraints and finite-entanglement revivals},\
  }\href {https://doi.org/10.1103/PhysRevB.101.024306} {\bibfield  {journal}
  {\bibinfo  {journal} {Phys. Rev. B}\ }\textbf {\bibinfo {volume} {101}},\
  \bibinfo {pages} {024306} (\bibinfo {year} {2020})}\BibitemShut {NoStop}%
\bibitem [{\citenamefont {Halimeh}\ \emph {et~al.}(2023)\citenamefont
  {Halimeh}, \citenamefont {Barbiero}, \citenamefont {Hauke}, \citenamefont
  {Grusdt},\ and\ \citenamefont {Bohrdt}}]{SHalimeh2023Quantum}%
  \BibitemOpen
  \bibfield  {author} {\bibinfo {author} {\bibfnamefont {J.~C.}\ \bibnamefont
  {Halimeh}}, \bibinfo {author} {\bibfnamefont {L.}~\bibnamefont {Barbiero}},
  \bibinfo {author} {\bibfnamefont {P.}~\bibnamefont {Hauke}}, \bibinfo
  {author} {\bibfnamefont {F.}~\bibnamefont {Grusdt}},\ and\ \bibinfo {author}
  {\bibfnamefont {A.}~\bibnamefont {Bohrdt}},\ }\bibfield  {title} {\bibinfo
  {title} {Robust quantum many-body scars in lattice gauge theories},\ }\href
  {https://doi.org/10.22331/q-2023-05-15-1004} {\bibfield  {journal} {\bibinfo
  {journal} {{Quantum}}\ }\textbf {\bibinfo {volume} {7}},\ \bibinfo {pages}
  {1004} (\bibinfo {year} {2023})}\BibitemShut {NoStop}%
\bibitem [{\citenamefont {Liu}\ \emph {et~al.}(2023)\citenamefont {Liu},
  \citenamefont {Lv}, \citenamefont {Yang},\ and\ \citenamefont
  {Zou}}]{SLiu2023CPL}%
  \BibitemOpen
  \bibfield  {author} {\bibinfo {author} {\bibfnamefont {Y.}~\bibnamefont
  {Liu}}, \bibinfo {author} {\bibfnamefont {S.}~\bibnamefont {Lv}}, \bibinfo
  {author} {\bibfnamefont {Y.}~\bibnamefont {Yang}},\ and\ \bibinfo {author}
  {\bibfnamefont {H.}~\bibnamefont {Zou}},\ }\bibfield  {title} {\bibinfo
  {title} {Signatures of quantum criticality in the complex inverse temperature
  plane},\ }\href@noop {} {\bibfield  {journal} {\bibinfo  {journal} {Chinese
  Physics Letters}\ }\textbf {\bibinfo {volume} {40}},\ \bibinfo {pages}
  {050502} (\bibinfo {year} {2023})}\BibitemShut {NoStop}%
\end{thebibliography}
%



\widetext
\begin{center}
\textbf{\large Supplementary information for `Detecting Many-Body Scars from Fisher Zeros'}
\end{center}

\setcounter{equation}{0}
\setcounter{figure}{0}
\setcounter{table}{0}
\makeatletter
\renewcommand{\thefigure}{S\arabic{figure}}
\renewcommand{\thetable}{S\arabic{table}}
\renewcommand{\theequation}{S\arabic{equation}}
\renewcommand{\bibnumfmt}[1]{[S#1]}
\renewcommand{\citenumfont}[1]{S#1}
\makeatother
\newcolumntype{C}{>{\centering\arraybackslash}X}

\section{Fisher zeros of the $\bar{P}X\bar{P}$ model and the Fixed-point line}

\begin{figure}[h]
    \includegraphics[width=0.9\textwidth]{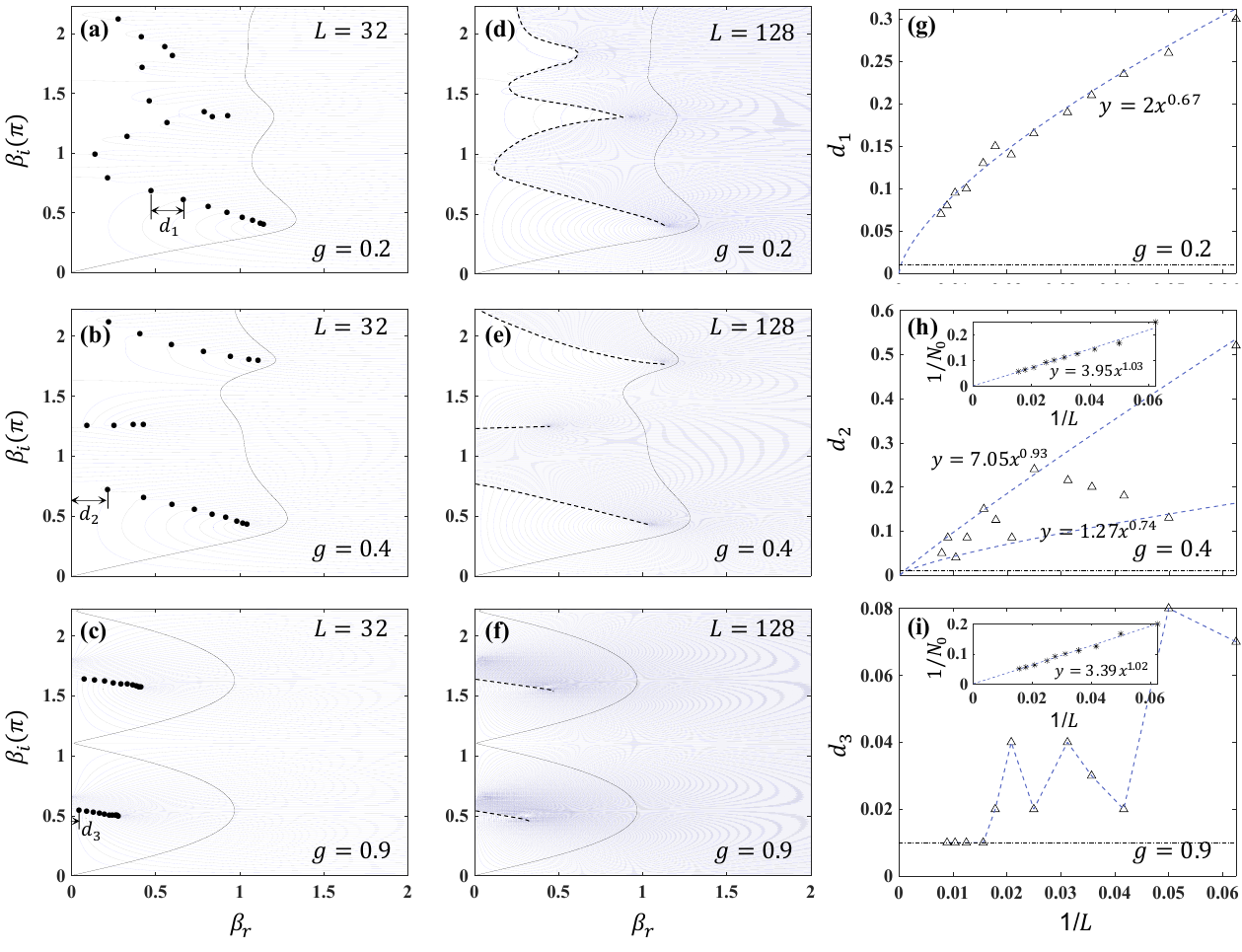}%
    \caption{Fisher zeros of the $\bar{P}X\bar{P}$ model for $g = 0.2$, 0.4, and 0.9, respectively. (a-c) Results at $L=32$ for different values of $g$: the blue and grey lines represent the solutions of $\mathrm{Re} Z=0$ and $\mathrm{Im}Z=0$, respectively, with their intersections indicating the Fisher zeros (black dots). The black line denotes the fixed-point line of the corresponding system. (d-f) The results for the same quantities at $L=128$. The Fisher zeros become denser and align along the dashed guide line. (g-i) Finite-size scaling results with the dashed black lines indicating a numerical precision of 0.01. (g) The difference $d_1$ in $\beta_r$ between the two selected Fisher zeros [shown in (a)] as a function of $1/L$; the dashed blue line represents the fit $y=2x^{0.67}$. (h,i) The real part $d_2$ or $d_3$ of the selected Fisher zeros [shown in (b) or (c)] exhibits oscillatory behavior as a function of $1/L$, with the upper and lower envelopes fitted by $y=7.05x^{0.93}$ and $y=1.27x^{0.74}$ in (h). The insets show the inverse of the number of Fisher zeros ($\sim 1/N_0$) in the lowest segment as a function of $1/L$, with the fitting functions shown as $y=3.95x^{1.03}$ in (h) and $y=3.39x^{1.02}$ in (i).}
    \label{fig:figS1}
\end{figure}

In the main text, we presented the Fisher zeros of the $\bar{P}X\bar{P}$ model in the complex $\beta$-plane for system size $L=64$. We further compute results for different system sizes. For example, Fig.~\ref{fig:figS1}(a–f) show the Fisher zeros at $L=32$ and $L=128$ for different values of $g$. As $L$ increases, the Fisher zeros become increasingly dense. More specifically, the finite-size scaling calculations [Fig.~\ref{fig:figS1}(g–i)] quantitatively verify that, in the thermodynamic limit, the Fisher zeros of the $\bar{P}X\bar{P}$ model form continuous lines. We first analyze the distance between two neighboring zeros or the real part of the zero closest to the $\beta_i$-axis as functions of $L$. For instance, at $g=0.2$, all Fisher zeros are off the $\beta_i$-axis, and the distance $d_1$ between two selected zeros decreases with $L$ following a power-law decay [Fig.~\ref{fig:figS1}(g)]. At $g=0.4$, the distance $d_2$ between the leftmost zero in the lowest segment and the $\beta_i$-axis decreases with increasing $L$, albeit with some oscillations. Power-law fits to the upper and lower envelopes confirm this overall decay [Fig.~\ref{fig:figS1}(h)]. These results suggest that the crossover region between QMBS and SBETH around $g\sim 0.3$ is weakly affected by system size. We perform similar calculations at $g=0.9$, where the distance $d_3$ from the leftmost zero to the $\beta_i$-axis becomes much smaller than that in the crossover regime. As $L$ increases, this distance approaches our numerical precision limit [Fig.~\ref{fig:figS1}(i)]. We further support our conclusions in the thermodynamic limit by analyzing the density of Fisher zeros. For example, at $g=0.4$ and $g=0.9$, the number of Fisher zeros $N_0$ in the lowest segment of the $\beta$-plane increases linearly with $L$.

\begin{figure}[h]
    \includegraphics[width=0.45\textwidth]{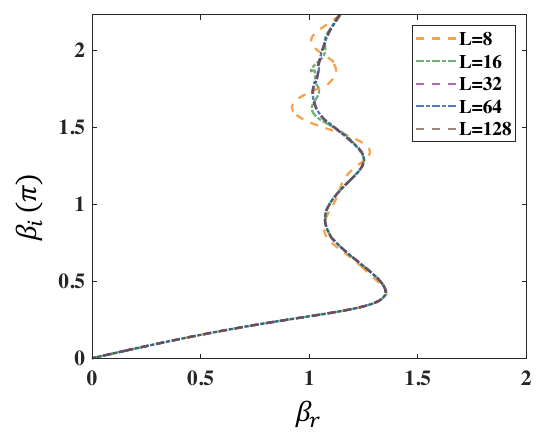}%
    \caption{The solutions of $S_0=1$ in the $\beta$-plane for the $\bar{P}X\bar{P}$ model at $g = 0$ at different system size $L$. As $L$ increases, these solutions converge to a single line, indicating that the condition $S_0=1$ can serve as a fixed point of the complex-$\beta$ RG flow.}
    \label{fig:figS2}
\end{figure}

In the main text, we defined $S_0=|Z(\beta_r,t)/Z(0,0)|^2$ [the spectral form factor normalized by $Z(0,0)$]. In Fig.~\ref{fig:figS1}(a-f), we also show the corresponding $S_0=1$ line for various system parameters $g$ and $L$.  The contour defined by $S_0=1$ in the complex-$\beta$ plane can be interpreted as a return rate of thermofield double (TFD) state of $Z(0,0)$, which is most intuitively visible in the case of $g=0.9$ [Fig.~\ref{fig:figS1}(c,f)]. There, the $S_0=1$ contour bends back toward the $\beta_i$-axis, suggesting the presence of Rabi-like oscillations of many states. In the intermediate $g$ regime, the structure of $S_0=1$ line in the complex plane shows no qualitative changes, indicating that the $S_0=1$ contour alone cannot pinpoint the precise location of the crossover between QMBS and SBETH. This motivates the need to incorporate additional information from the Fisher zeros.

From the perspective of renormalization group (RG) analysis, the value of $S_0$ can be used to construct the RG flow~\cite{Liu2024PRR}, and the $S_0=1$ contour can also be viewed as the extension of the high temperature RG fixed point at $\beta=0$ into the complex-$\beta$ plane, forming what we refer to as a ``fixed-point line". This interpretation is supported by the results shown in Fig.~\ref{fig:figS2}, where we observe that the $S_0=1$ contour converges to a line as $L$ increases.
In general, we can define a RG transformation for complex $\beta$ with a renormalization scale $\mu=b/L$, where $b$ is the rescaling factor, and the generalized Symanzik $\mathscr{B}(\beta)$ function as 
\begin{equation}
\mathscr{B}(\beta)=\frac{d\beta}{d\ln\mu}.
\end{equation}
In the large $L$ limit, since $\beta$ does not vary with $\mu$ (or $L$) along the $S_0=1$ contour line, we have $\mathscr{B}(\beta)=0$ on this line, that it is marginal with respect to the fixed point at $\beta=0$. We employ the two-lattice matching method~\cite{RGflow2010,Zou2011PRD} and set the rescaling factor $b=2$, which iteratively compares the values of $S_0$ between systems of size $L/2$ and $L$, to construct the RG flow in the complex-$\beta$ plane from the Fisher zeros (or the large-$\beta$ region) toward the $S_0=1$ line. 
Specifically, we can start from an arbitrary $\beta_i$, solve the equation 
\begin{equation}
S_0(\beta_{i+1},L)=S_0(\beta_i,L/2), 
\label{eq:matching}
\end{equation}
 to obtain $\beta_{i+1}$, and then use $\beta_{i+1}$ as the new starting point for iterative computation, continuing this process until $\beta$ reaches the fixed line $S_0=1$. 
 Note that for the same $\beta_i$, all solutions $\beta_{i+1}$ satisfying Eq.~\ref{eq:matching} actually form a line. We choose the $\beta_{i+1}$ that is closest to $\beta_i$.
In the main text, we presented the RG flow results for the matching between $L=32$ and $L=64$. 
In contrast to Ref.~\cite{RGflow2010,Zou2011PRD}, where the compare-$\beta$ RG flow terminates at an unique fix point ($\beta=0$), the newly defined RG flow here terminates on an extended fixed-point line . The flow toward Fisher zeros in Ref.~\cite{RGflow2010,Zou2011PRD} exhibits ambiguous behavior for finite systems, whereas the newly defined RG flow always flows outward from the Fisher zeros, thereby avoiding this issue.

\begin{figure}[h]
    \includegraphics[width=0.9\textwidth]{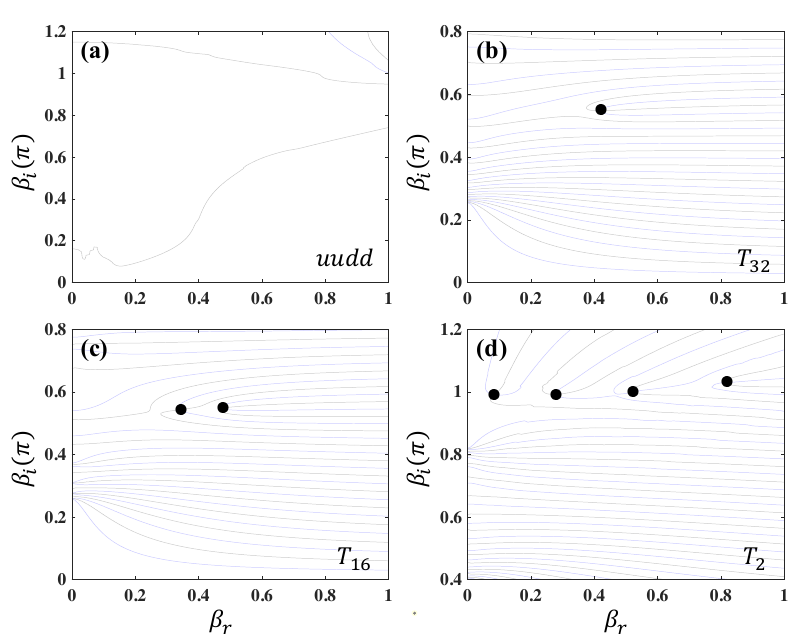}%
    \caption{The Fisher zeros of boundary partition function $Z_i$ for different states in the $\bar{P}X\bar{P}$ model at $g=0$ (the $PXP$ model) with size $L=32$: (a) $uudd$ state, (b) $T_{32}$ state, (c) $T_{16}$ state, and (d) $T_2$ state (or AF state). The markings of lines and points are the same as in Fig. 2 of the main text. In (a), there are no zeros in the considered region, while (b)-(d) contain 1, 2, and 4 zeros, respectively. Their positions and the trend of approaching the $\beta_i$ axis differ significantly.}
    \label{fig:figS4}
\end{figure}

\section{Fisher zeros of $Z_i$ in the $\bar{P}X\bar{P}$ model}

In the $\bar{P}X\bar{P}$ model, QMBS and SBETH not only exhibit different configurations of the Fisher zeros of $Z$ but also demonstrate distinct properties in the Fisher zeros of $Z_i$. At $g=1$, the zeros of all the $Z_i$ coincide with those of $Z$. As $g$ decreases and the blockade interaction is introduced, the zeros of most $Z_i$ remain close to the same position.
This is also reflected in the dense lines of the solutions of $\mathrm{Re}Z=0$ and $\mathrm{Im}Z=0$ in Fig.~2(a) of the main text. In contrast, in the QMBS region, $Z$ does not have Fisher zeros on the $\beta_i$-axis, implying that the zeros of $Z_i$ on the $\beta_i$ axis (if they exist) cannot coincide at the same position. Consequently, the dynamical behavior of each state will be completely different. We verify this conclusion by calculating the zeros of $Z_i$ for different states at $g=0$ with $L=32$. For a product state whose spin is configured with an up-up-down-down ($uudd$) pattern, we find that there are no zeros in a large region near the $\beta_r$ and $\beta_i$ axes [Fig.~\ref{fig:figS4}(a)]. 
We further consider different $T_i$ states (defined by a repeating pattern $uddd\cdots$, i.e. 
an up spin followed by $i-1$ down spins). For states closer to the FM state, we find that there are only one and two zeros exists near the axes in the $T_{32}$ and $T_{16}$ state, respectively [Fig.~\ref{fig:figS4}(b)(c)]. In contrast, for the scar state $T_2$ (or AF state), its zeros exhibit a clear tendency to approach the $\beta_i$ axis, and the positions differ significantly from those of the $T_{32}$ and $T_{16}$ states [Fig.~\ref{fig:figS4}(d)]. From this perspective, the differences in the zeros of $Z_i$ for all possible states in the Hilbert space are responsible for the zeros of the total $Z$ moving away from the $\beta_i$-axis.

\begin{figure}[h]
    \includegraphics[width=0.9\textwidth]{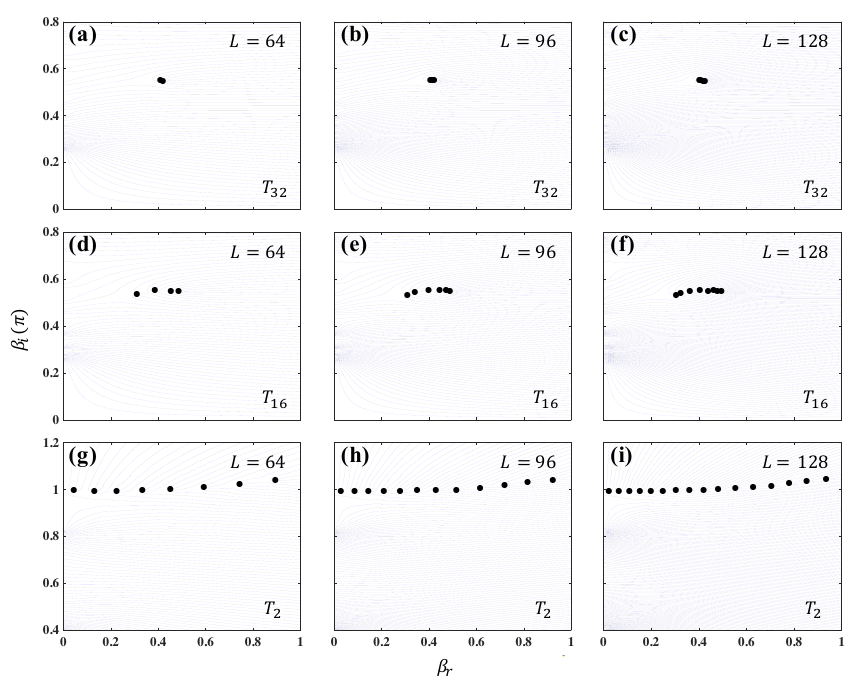}%
    \caption{The Fisher zeros of $Z_i$ for different states ($T_{32}$, $T_{16}$, and $T_2$) in $\bar{P}X\bar{P}$ model at $g = 0$ with larger system sizes $L=64$, 96, and 128. The number of Fisher zeros shows a linear growth trend with increasing $L$. (a-f) For the non-scar state $T_{32}$ and $T_{16}$, the Fisher zeros marked off the $\beta_i$-axis in Fig.~\ref{fig:figS4} still do not approach the $\beta_i$-axis even as $L$ increases. (g-i) In contrast, for the scar state $T_2$, the marked zeros in Fig.~\ref{fig:figS4} gradually approach the $\beta_i$-axis as $L$ increases.}
    \label{fig:figS5}
\end{figure}

To verify the above conclusions regarding the distinct behaviors of Fisher zeros for different $Z_i$, we further calculate the Fisher zeros for larger system sizes. As $L$ increases, similar to the results in Fig.~\ref{fig:figS1}, the number of Fisher zeros grows linearly with $L$ for all the cases of initial states, indicating that in the thermodynamic limit, the zeros also form continuous lines. However, the line configurations formed by non-scar and scar initial states are significantly different. For the $T_{32}$ and $T_{16}$ states, the marked zeros in Fig.~\ref{fig:figS4} tend to approach a segment that does not touch the $\beta_i$-axis in the thermodynamic limit [Fig.~\ref{fig:figS5}(a-f)]. In contrast, for the scar state $T_2$, the marked zeros in Fig.~\ref{fig:figS4} move closer to the $\beta_i$-axis as $L$ increases [Fig.~\ref{fig:figS5}(g-i)].

\section{Fisher zeros of a lattice gauge field related model}

To verify the general relationship between QMBS and the configuration of Fisher zeros in the complex-$\beta$ plane, we further compute the Fisher zeros of the partition function for a cluster spin model closely related to a lattice gauge theory. Specifically, we start from the 1+1D lattice field theory model
\begin{equation}
H=t\sum_{i}(c_i^{\dagger}\sigma_{i,i+1}^zc_{i+1}+\mathrm{h.c})+h\sum_{i}\sigma_{i,i+1}^x,
\label{eq:modelLGSo}
\end{equation}
where $c^\dagger_i$ and $c_i$ denote the creation and annihilation operators for hard-core bosons at lattice site $i$, and $\sigma^{x/z}_{i,i+1}$ represents the gauge field on the links connecting neighboring lattice sites. This model can be reformulated using gauge-invariant local spin operator as~\cite{Sergej2020prl}:
\begin{equation}
H=\frac{t}{2}\sum_i(Z_{i,i+1}-X_{i-1,i}Z_{i,i+1}X_{i+1,i+2})+h\sum_i X_{i,i+1},
\end{equation}
where $X_{i,i+1}=\sigma_{i,i+1}^x$ and $Z_{i,i+1}=(c_i^\dagger-c_i)\sigma_{i,i+1}^z(c_i^{\dagger}+c_i)$ are gauge-invariant local operators. By defining the gauge link variables as spin operators on lattice sites and applying a spin-basis rotation, the spin model can be rewritten in the following form~\cite{Iadecola2020prb}:
\begin{equation}
   {H} = \frac{t}{2}\sum_{i=1}^{L}(\sigma^x_{i}-\sigma^z_{i-1}\sigma^x_{i}\sigma^z_{i+1})+h\sum_i\sigma^z_{i}.
    \label{eq:modelLGS}
\end{equation}

\begin{figure}[h]
    \includegraphics[width=0.9\textwidth]{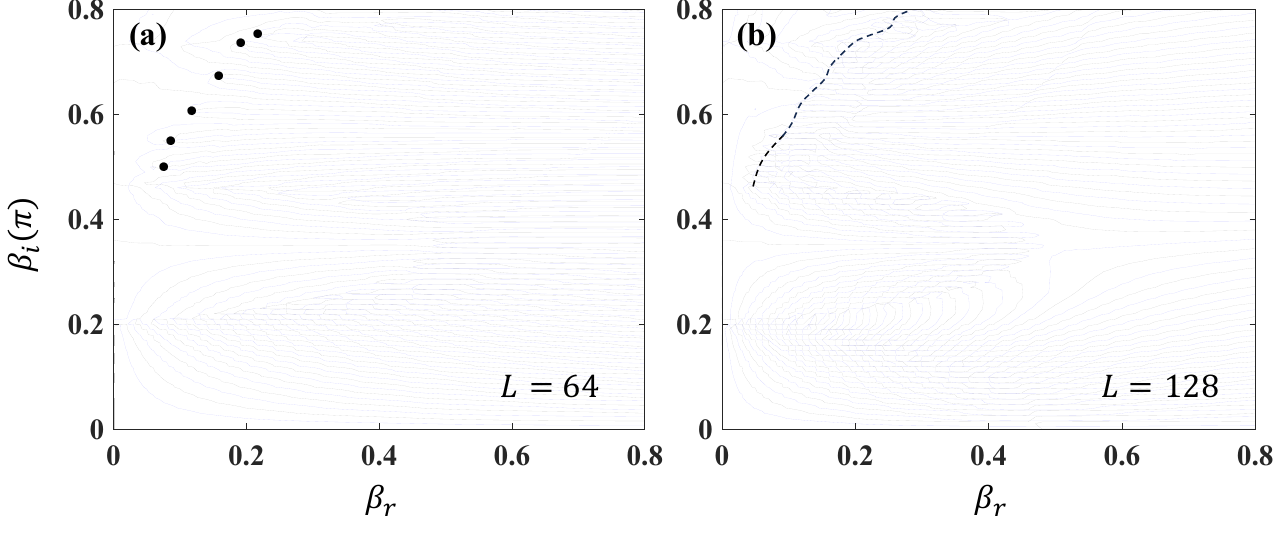}%
    \caption{Fisher zeros of the cluster spin model defined in Eq.~\ref{eq:modelLGS}: (a) for $L=64$, (b) for $L=128$, with $t=2$ and $h=0.1$. The zeros off the $\beta_i$-axis are marked with black dots or dashed line.}
    \label{fig:figS6}
\end{figure}

\begin{figure}[h]
    \includegraphics[width=0.45\textwidth]{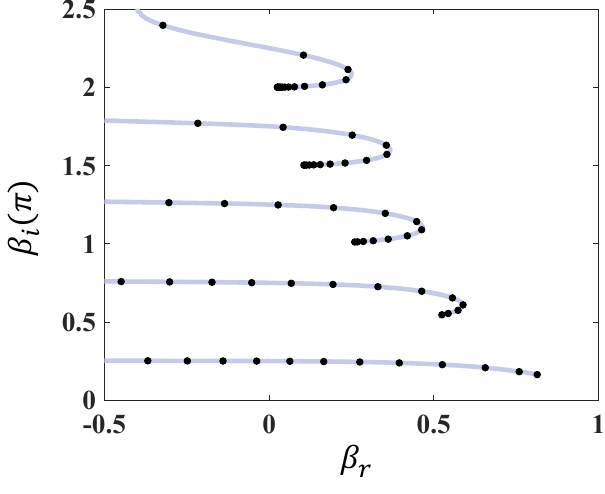}%
    \caption{Fisher zeros of the Ising model with $h_l = 0.5$ in the finite system and in the thermodynamic limit. The dots represent the zeros for $L = 32$, and the line indicates the zeros in the thermodynamic limit.}
    \label{fig:figS7}
\end{figure}

All three models described by Eqs.~\ref{eq:modelLGSo}-\ref{eq:modelLGS} support scar states~\cite{Iadecola2020prb,Halimeh2023Quantum}. We compute the partition function and Fisher zeros for the spin model Eq.~\ref{eq:modelLGS}, and the results (Fig.~\ref{fig:figS6}) show similarities to those of the Ising model with fields discussed in the main text. In particular, the Fisher zero lines associated with quantum many-body scars, which deviate from the $\beta_i$-axis, coexist with other zero configurations. This reveals a more complex structure compared to the pure scar case in the $PXP$ model. Although both the $PXP$ model and this spin model contain a cluster Ising term, the competition with different additional terms results in distinct Fisher zero patterns.

\section{details on the Ising model with external fields}

When the Ising model with external fields in the main text contains only a longitudinal field $h_l$, the partition function $Z$ of the system can be solved exactly, with the solution given by
\begin{equation}
   Z=e^{\beta}\cosh^L\beta h_l(\lambda_+^L+\lambda_-^L),
    \label{eq:Zising}
\end{equation}
where $\lambda_\pm=1\pm\tanh\beta h_l\sqrt{1+e^{-4\beta}/\sinh^2\beta h_l}$. Similar to the exact Fisher zero solution of the transverse-field Ising model~\cite{Liu2023CPL}, the zeros consist of two parts: one set lies on the $\beta_i$-axis, originating from the $\cosh\beta h_l$ term; the other forms a continuous line in the thermodynamic limit, with the solution given by
\begin{equation}
   \ln\left|\frac{\lambda_+}{\lambda_-}\right|^2=0.
    \label{eq:solutionZ}
\end{equation}
We find that the off–$\beta_i$-axis Fisher zeros in the finite-size system still lie on the continuous line of zeros obtained analytically in the thermodynamic limit (Fig.~\ref{fig:figS7}), further supporting the validity of the results presented in the main text.

\begin{figure}[h]
    \includegraphics[width=0.9\textwidth]{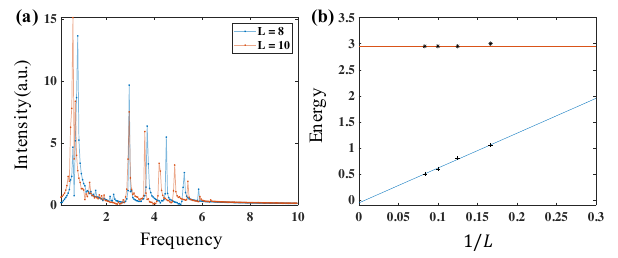}%
    \caption{Finite size scaling of the spectrum peak from ED calculation. (a) The dynamic spectrum of the FM state in the system with $L=8$ and $L=10$. (b) The scaling of the first frequency peak and the scar-related peak with system size $L$.}
    \label{fig:figS8}
\end{figure}

To demonstrate the stability of the scar-related frequency peak, we perform a finite-size scaling analysis based on the ED results discussed in the main text. By varying $L$, we found that all peaks except the one at frequency 2.9 decrease as $L$ increases, exhibiting an inverse relationship with system size $L$ (Fig.~\ref{fig:figS8}).

\end{document}